\documentclass[a4paper,11pt]{article}
\pdfoutput=1

\usepackage{jheppub}

\usepackage{amsmath,color,amssymb}
\usepackage{graphicx}
\usepackage{hyperref}

\newcommand{\drawsquare}[2]{\hbox{%
\rule{#2pt}{#1pt}\hskip-#2pt
\rule{#1pt}{#2pt}\hskip-#1pt
\rule[#1pt]{#1pt}{#2pt}}\rule[#1pt]{#2pt}{#2pt}\hskip-#2pt
\rule{#2pt}{#1pt}}
\newcommand{\Yfund}{\raisebox{-.5pt}{\drawsquare{6.5}{0.4}}}
\newcommand{\Ysymm}{\raisebox{-.5pt}{\drawsquare{6.5}{0.4}}\hskip-0.4pt%
        \raisebox{-.5pt}{\drawsquare{6.5}{0.4}}}
\newcommand{\Yasymm}{\raisebox{-3.5pt}{\drawsquare{6.5}{0.4}}\hskip-6.9pt%
        \raisebox{3pt}{\drawsquare{6.5}{0.4}}}

\newcommand{\lsim}{\!\mathrel{\hbox{\rlap{\lower.55ex \hbox{$\sim$}} \kern-.34em \raise.4ex \hbox{$<$}}}}
\newcommand{\gsim}{\!\mathrel{\hbox{\rlap{\lower.55ex \hbox{$\sim$}} \kern-.34em \raise.4ex \hbox{$>$}}}}
\newcommand{\Tr}{\text{ Tr}}
\newcommand{\tr}{\text{ tr}}

\newcommand{\beq}{\begin{eqnarray}}
\newcommand{\eeq}{\end{eqnarray}}
\newcommand{\nn}{\nonumber}

\title{SIMP Spectroscopy}

\author{Yonit Hochberg,${}^{1,2}$}\emailAdd{yonit.hochberg@berkeley.edu}
\author{Eric Kuflik,${}^3$}\emailAdd{kuflik@cornell.edu}
\author{Hitoshi Murayama,${}^{1,2,4,5}$}\emailAdd{hitoshi@berkeley.edu}\emailAdd{hitoshi.murayama@ipmu.jp}
\affiliation{${}^1$Ernest Orlando Lawrence Berkeley National Laboratory, University of California, Berkeley, CA 94720, USA}
\affiliation{${}^2$Department of Physics, University of California, Berkeley, CA 94720, USA}
\affiliation{${}^3$Department of Physics, LEPP, Cornell University, Ithaca NY 14853, USA}
\affiliation{${}^4$Kavli Institute for the Physics and Mathematics of the
  Universe (WPI), University of Tokyo Institutes for Advanced Study, University of Tokyo,
  Kashiwa 277-8583, Japan}
\affiliation{${}^5$Center for Japanese Studies, University of California, Berkeley, CA 94720, USA}

\preprint{UCB-PTH 15/16, IPMU15-0218}

\abstract{We study the interactions between strongly interacting massive particle dark matter and the Standard Model via a massive vector boson
that is kinetically mixed with the hypercharge gauge boson. The relic abundance is set by $3\to2$ self-interactions of the   dark matter, while
the interactions with the vector mediator  enable kinetic equilibrium between the dark and visible sectors. We show that a wide range of
parameters is phenomenologically viable  and can be probed in various ways. Astrophysical and cosmological  constraints are  evaded due to the
$p$-wave nature of dark  matter annihilation into visible particles, while direct detection  methods using electron recoils can be sensitive to
parts of the  parameter space. In addition, we propose performing spectroscopy of  the strongly coupled dark sector at $e^+ e^-$ colliders, where
the  energy of a mono-photon can track the resonance structure of the  dark sector. Alternatively, some resonances may decay back into  Standard
Model leptons or jets, realizing `hidden valley'  phenomenology at the LHC and ILC in a concrete fashion.
}
\begin{document}
\maketitle
\flushbottom

\section{Introduction}\label{sec:intro}
The weakly interacting massive particle (WIMP) has long been the dominant paradigm for explaining the presence of dark matter. This is because of
the so-called `WIMP miracle'---that weak-scale dark matter with weak interactions gives the correct relic abundance if the dark matter annihilates
to Standard Model (SM) particles---and that we expect new weak-scale particles to exist and address the hierarchy problem. There is, however, no
compelling evidence of the WIMP's existence, and new particles at the weak-scale have yet to be discovered.  As a result, considerable theoretical
and experimental effort has been put into exploring dark-matter scenarios beyond the WIMP.

One recent proposal to explain the particle nature of dark matter is the strongly interacting massive particle (SIMP)
mechanism~\cite{Hochberg:2014dra}. SIMPs have several distinguishing characteristics compared to WIMPs. First, the process responsible for the
dark matter relic abundance is a $3\to2$ self-annihilation process consisting entirely of dark matter particles.  This mechanism of thermal
freezeout predicts dark matter that is lighter than the WIMP, with masses typically between an MeV and a GeV. Indeed, QCD-scale dark matter with
strong self-couplings gives the correct relic abundance. The strong self-interactions of the SIMP also lead to strong
self-scatterings, which can address longstanding puzzles in small-scale structure. The $3\to2$ process pumps heat into the system~\cite{Carlson:1992fn}, and so the dark
matter must dump its entropy, in order to stay cool during freezeout. This can be done by interacting either with new light degrees of freedom or
with abundant light SM particles. In the latter case, couplings to the SM of a particular size are necessary for maintaining thermal equilibrium
with the thermal bath, leading to distinct experimental dark-matter signatures.

In Ref.~\cite{Hochberg:2014kqa} it was proposed that SIMPs could be the pions (pseudo-Nambu-Goldstone bosons) of a strongly coupled QCD-like
confining hidden sector, where the Wess-Zumino-Witten term~\cite{Wess:1971yu,Witten:1983tw,Witten:1983tx} generated the $3\to2$ number-changing
process.  Requiring the validity of chiral perturbation theory, together with dark matter self-interaction constraints, suggests that the mass of
the pion and its decay constant should be of order a few~$\times~100$~MeV, resembling the pions and kaons of the SM.

Various aspects of SIMP dark matter are being studied in the literature (see, {\it e.g.},
Refs.~\cite{Choi:2015bya, Lee:2015gsa, Bernal:2015bla, Schwaller:2015tja, Bernal:2015xba, Bernal:2015ova, Hansen:2015yaa}). In Ref.~\cite{Lee:2015gsa},
a particular mediator between the SM and SIMPs was studied, towards maintaining thermal contact between the two sectors. The setup was of a
$U(1)_D$ symmetry under which the pions are charged, and whose gauge boson kinetically mixes with the hypercharge gauge boson of the SM.

In this work, we
explore in further detail the prospects for kinetically mixed vector mediation between the SIMP and SM sectors. We include the possibility that the dark matter belongs to various different confining gauge groups, finding appropriate subgroups of the residual global symmetries that can be gauged. We find the experimental constraints on the kinetically mixed parameter space, and give future experimental probes of these interactions.

Importantly, we propose performing \emph{spectroscopy} of the strongly coupled sector at $e^+
e^-$ colliders, where the energy of a mono-photon, in  $e^+ e^-\to \gamma + \rm inv$,  can be used as a tracer for the resonance structure of the
SIMP sector. More generally,  dark spectroscopy can be done for any strongly coupled dark sector that couples (in a conceptually similar way) to
the SM.

The outline of this paper is as follows. Section~\ref{sec:SIMP} reviews the pion-setup of the SIMP mechanism, and Section~\ref{sec:darkphoton}
sets up the kinetically mixed hidden photon. In Section~\ref{sec:simpU1} we describe the dark photon as a mediator between SIMPs and the SM: this
includes the embedding of the dark photon into the SIMP setup for different confining gauge groups, as well as thermalization and annihilation.
Section~\ref{sec:decays} details the decays of the dark sector, visible and invisible. We describe our proposal of dark spectroscopy in
Section~\ref{sec:spec}, discuss future prospects in Section~\ref{sec:future}, and conclude in Section~\ref{sec:conc}.

\section{SIMP review}\label{sec:SIMP}

We begin by reviewing the SIMP setup presented in Ref.~\cite{Hochberg:2014kqa}, where the dark matter is a pion (pseudo-Nambu-Goldstone boson) of
a confining hidden sector, with the Wess-Zumino-Witten (WZW) action generating the $3\to 2$ process.

We consider an $SU(N_c$) gauge theory with $N_f$ Dirac fermions, an $SO(N_c$) gauge theory with $N_f$ Weyl fermions, or an $Sp(N_c$) gauge theory
with $2N_f$ Weyl-fermions, $q_i$, in the $N_c$-dimensional representation of the gauge group, with $N_c$ the number of colors ($N_c$ is even for
$Sp(N_c)$).  Neglecting quark masses, the Lagrangian takes the simple form
\beq\label{eq:L}
{\cal L} = -\frac{1}{4} F_{\mu\nu}^a
F^{\mu\nu a} + {q}^\dagger_i i \bar{\sigma}^\mu D_\mu q_i \,.
\eeq
The Lagrangian contains an exact global-symmetry, $G$, which with a sufficiently small $N_f$ for a given $N_c$, leads to chiral symmetry
breaking of the global symmetry down to a subgroup $H$. The expected symmetry-breaking pattern $G/H$ is $SU(N_f)\times SU(N_f)/SU(N_f)$,
$SU(N_f)/SO(N_f$), and $SU(2 N_f)/Sp(2 N_f$), for an $SU(N_c)$, $SO(N_c)$, and $Sp(N_c)$ gauge theory, respectively.

The low-energy effective chiral action for the above theories is
\beq
S_{\text{eff}}&=& \int{} d^4 x \frac{f_\pi^2}{16} \Tr \partial_\mu
\Sigma \partial^\mu \Sigma^\dagger+ \Gamma_{\rm WZW}\,,
\eeq
where $\Sigma$ parameterizes the coset space $G/H$\,,
\begin{equation}
\label{eq:sigma}
  \Sigma  = \left\{ \begin{array}{lcc}
  \exp( 2i \pi/f_\pi) I & &  \quad {\rm for~}SU(N_c)\ {\rm or}\ SO(N_c)\,,\\
  \exp( 2i \pi/f_\pi) J  & &  \quad {\rm for~} Sp(N_c)\,.
  \end{array} \right.
\end{equation}
Here $\pi = \pi_a T_a $,  $T_a$ are the broken generators of $G/H$ and $J = i\sigma_2 \otimes I_{N_f}$ is the $Sp(2 N_f$) invariant.  We use the
 normalization ${\rm Tr} T^a T^b = 2 \delta^{ab}$.

 The ungauged Wess-Zumino-Witten  action, $\Gamma_{\rm WZW}$,  takes the form:
 \beq
 \Gamma_{\rm WZW} = \frac{-i N_c }{240 \pi^2}  \int_{M^5} \Tr(d\Sigma \Sigma^{-1})^5\,,
 \eeq
 where the integral is over a five-dimensional manifold ($M^5$) whose boundary is identified  with the ordinary four-dimensional Minkowski space.
 The WZW-action is non-vanishing provided that the fifth homotopy group of the coset  space, $\pi_5 (G/H)$, is non-trivial. This is satisfied when
 $N_f \ge 3$ for $SU(N_c$) and $SO(N_c$) gauge theories and  $N_f \ge 2$ for $Sp(N_c$) gauge theories.

 Expanding the effective-action to leading order in the pion fields, the interaction Lagrangian is
\begin{equation}\label{eq:wzw}
  {\cal L}_{\rm int}
  = -\frac{1}{6 f_\pi^2} \Tr\left(\pi^2 \partial^{\mu} \pi \partial_{\mu}  \pi  -\pi \partial^{\mu} \pi \pi \partial_{\mu} \pi \right)+
  \frac{2N_c}{15 \pi^2 f_\pi^5}\epsilon^{\mu\nu\rho\sigma} \Tr \left(\pi \partial_\mu
  \pi \partial_\nu \pi \partial_\rho \pi \partial_\sigma \pi \right) +{\cal O}(\pi^6)\,.
\end{equation}
The 5-point interactions coming from the WZW-action, which involve five different pion fields, enable the $3\to 2$ process of the SIMP
mechanism.

The SIMPlest model is that of the smallest particle content, which is the $Sp(2)\simeq SU(2)$ gauge group with $N_f=2$, where the coset space is
$SU(4)/Sp(4) = SO(6)/SO(5) = S^5$, with  $\pi_5 (S^5) = {\mathbb Z}$.  The five pions are the minimum number of pions needed for the $3\rightarrow
2$ process via the WZW term. In what follows, this model will often be used as an example to demonstrate the phenomenology of the proposal here.

Without an explicit mass term for the quarks, the pions are exact Goldstone-bosons, and therefore massless. For the gauge theories considered
here, an $H$-invariant quark mass can be added to the Lagrangian,
\begin{equation}\label{eq:qmass}
 {\cal L}_{\rm mass}=
 \begin{cases}
      m_q I^{ij} \bar{q}_i q_j + h.c. \,, &   {\rm for~}SU(N_c) \\
          \frac{1}{2}  m_q I^{ij} {q}_i q_j + h.c.\,, &   {\rm for~}SO(N_c) \\
        \frac{1}{2}   m_q J^{ij} {q}_i q_j + h.c.\,, & {\rm for~}Sp(N_c)
   \end{cases}
\end{equation}
which induces a mass for all the pions\,,
\beq \label{eq:pimass}
 {\cal L}_{\rm eff-mass} &=&  \begin{cases}
      -4 f_\pi^2 m_\pi^2 \Tr \Sigma + h.c. &   {\rm for~}SU(N_c),~SO(N_c)  \\
       -4 f_\pi^2 m_\pi^2  \Tr J \Sigma + h.c. &   {\rm for~}Sp(N_c)
   \end{cases}\nn\\
   &= & -\frac{m_\pi^2}{4} \Tr  \pi^2 + \frac{m_\pi^2}{12 f_\pi^2}  \Tr \pi^4 +{\cal O}(\pi^6)\,.
\eeq
The $H$-invariant mass-term ensures that the chiral Lagrangian respects the unbroken symmetry. The pions transform as a non-trivial representation
under $H$ and thus are stable, as they are the lightest fields with non-trivial quantum numbers.  In what follows, we will always assume a mass-term for the quarks that
preserves some subgroup of the flavor symmetry $H$, which acts to stabilize the pions.

The thermally averaged $3 \to 2 $ annihilation cross-section can be calculated from the effective 5-point interaction in Eq.~\eqref{eq:wzw},
\beq\label{eq:xsec32}
\langle \sigma v^2\rangle_{3\to2} = \frac{5\sqrt{5}}{2\pi^5 x_f^2}\;\frac{N_c^2 m_\pi^5}{f_\pi^{10}}\; \frac{t^2}{N_\pi^3}\,,
\eeq
where $x_f=m_\pi/T_f \simeq 20$ and $T_f$ is the bath temperature at freeze-out.
The factor of $t^2/N_\pi^3$ is combinatorial factor that depends on $N_f$ and the choice of the confining gauge-theory; further details are found
in the Appendix of Ref.~\cite{Hochberg:2014kqa}. For a given $N_f$ and $N_c$, obtaining the correct relic abundance via $3 \to 2 $ annihilation
gives an explicit relationship between $m_\pi$ and $f_\pi$.

There are strong constraints on large dark-matter self scatterings from bullet-cluster
obsersvations~\cite{Clowe:2003tk,Markevitch:2003at,Randall:2007ph} and halo shape simulations~\cite{Rocha:2012jg,Peter:2012jh}, which give roughly
${\sigma_{\rm scatter}}/{m_{\rm DM}} \lesssim 1~{\rm cm}^2/{\rm g}$. In terms of the pions, the self-scattering cross section can be calculated to
leading order using the 4-point interactions of the chiral Langrangian [Eqs.~(\ref{eq:wzw}) and (\ref{eq:pimass})],
\beq\label{eq:xsec22}
\sigma_{\rm scatter}=\frac{m_\pi^2}{32 \pi f_\pi^4}\; \frac{a^2}{N_\pi^2}\,,
\eeq
where the factor of $a^2/N_\pi^2$ is a combinatorial factor that depends on $N_f$ and the choice of the confining gauge-theory; further details
can be  found in the Appendix of Ref.~\cite{Hochberg:2014kqa}.

Applying the constraints on self-interactions, that the pions produce the observed dark-matter relic abundance, and assuming validity of chiral
perturbation theory $m_\pi/f_\pi \lesssim 2\pi$, points to a preferred region of parameter space which is very similar to that of QCD:
\beq\label{eq:mpi}
m_\pi \sim 300~{\rm MeV}, ~~~~~~~~~~~~~ f_\pi \sim  {\rm few} \times m_\pi.
\eeq
This corresponds to the strongly interacting regime of the theory, where the strong dynamics can induce ${\cal O}(1)$ changes to the above;
Eq.~\eqref{eq:mpi} should be thought of as a mere proxy to the scales involved.

We note that the self-scattering cross sections predicted by Eq.~(\ref{eq:xsec22}) are of the right size to reconcile discrepancies in N-body simulations with the observed small-scale structure, such as the `core vs. cusp' and `two big to fail' puzzles~\cite{Spergel:1999mh,deBlok:2009sp,BoylanKolchin:2011de,Kaplinghat:2015aga,Zavala:2012us,Vogelsberger:2012ku,Rocha:2012jg,Peter:2012jh}. Additionally, a recent study of the Abell 3827 galaxy cluster
showed evidence for self-interactions of dark matter~\cite{Massey:2015dkw,Kahlhoefer:2015vua}.
Ref.~\cite{Massey:2015dkw} claimed the self-scattering cross section
should be $\sigma_{\rm scatter}/m = (1.7 \pm 0.7) \times 10^{-4}~{\rm cm}^2{\rm g}^{-1} \times
(t_{\rm infall}/10^9~{\rm yr})^{-2}$, but
a reanalysis in Ref.~\cite{Kahlhoefer:2015vua} finds
$\sigma_{\rm scatter}/m = 1.5~{\rm cm}^2 {\rm g}^{-1}$ as the preferred value, similar to the rate needed to address the small-scale structure puzzles and the rate expected from the SIMP mechanism.

\section{Dark photon review}\label{sec:darkphoton}

We consider a massive dark photon, which is the $U(1)_D$ gauge boson, that is kinetically mixed with the hypercharge gauge boson. Extensively
studied in the literature, here we summarize the relevant parts of the dark photon Lagrangian as well as relevant experimental limits, including
several new limits.

\subsection{Kinetic mixing}\label{ssec:mix}

We take a $U(1)_D$ gauge theory with a gauge field ${\cal A}_\mu$ which has kinetic mixing with the $U(1)_Y$ gauge field $B_\mu$,
\begin{equation}
  {\cal L}_{\cal A} =  -\frac{1}{4}  {\cal A}_{\mu\nu} {\cal A}^{\mu\nu}  - \frac{\sin\chi}{2} B_{\mu\nu} {\cal A}^{\mu\nu} +      \frac{1}{2} m_V^2 {\cal A}_\mu {\cal A}^\mu\, .
\end{equation}
The gauge boson's mass may arise either from a hidden-photon Higgs mechanism or the St\"uckelberg trick.

After electroweak symmetry breaking occurs, ${\cal A}_\mu$ becomes a mixture of the $Z$-boson and the dark photon
$V$~\cite{Hook:2010tw,Lee:2015gsa},
\begin{equation}
  V^\mu = - \frac{s_\zeta}{c_\chi} Z^\mu + \frac{c_\zeta}{c_\chi}
  {\cal A}^\mu\, ,
\end{equation}
where
\begin{equation}
  \tan 2\zeta
  = \frac{m_Z^2 s_W \sin 2\chi}{m_V^2 - m_Z^2 (c_\chi^2 - s_W^2 s_\chi^2)}\, .
\end{equation}
The mass eigenvalues are then
\begin{equation}
  m^2_\pm = \frac{1}{2} \left[
    m_Z^2 (1+s_W^2 t_\chi^2) + \frac{m_V^2}{c_\chi^2}
    \pm \sqrt{ \left(m_Z^2 (1+s_W^2 t_\chi^2) + \frac{m_V^2}{c_\chi^2}
    \right)^2 - \frac{4}{c_\chi^2} m_Z^2 m_V^2 }\  \right]\,.
\end{equation}
The vector bosons couple to the current operators as
\begin{eqnarray}
  {\cal L}_D =& &A_\mu J_{EM}^\mu
  + Z_\mu \left[c_W s_\zeta t_\chi J_{EM}^\mu + (c_\zeta - s_W t_\chi s_\zeta) J_Z^\mu
  - \frac{s_\zeta}{c_\chi} J_D^\mu \right] \nonumber \\
  & &
  + V_{\mu} \left[ J_{EM}^\mu (-c_W c_\zeta t_\chi)
  + J_Z^\mu (s_\zeta + s_W t_\chi c_\zeta) + \frac{c_\zeta}{c_\chi}
      J_D^\mu \right]\,,
      \label{eq:LD}
\end{eqnarray}
For later convenience, we define the coupling strength of the dark photon to the electromagnetic and $Z$ currents,
\beq
\epsilon_\gamma \equiv -c_W c_\zeta t_\chi\quad {\rm and} \quad \epsilon_Z \equiv s_\zeta + s_W t_\chi c_\zeta\,.
\eeq
%

\subsection{Experimental limits on dark photons}\label{ssec:expU1}

\begin{figure}[t!]
  \centering
  \includegraphics[clip,width=0.49\textwidth]{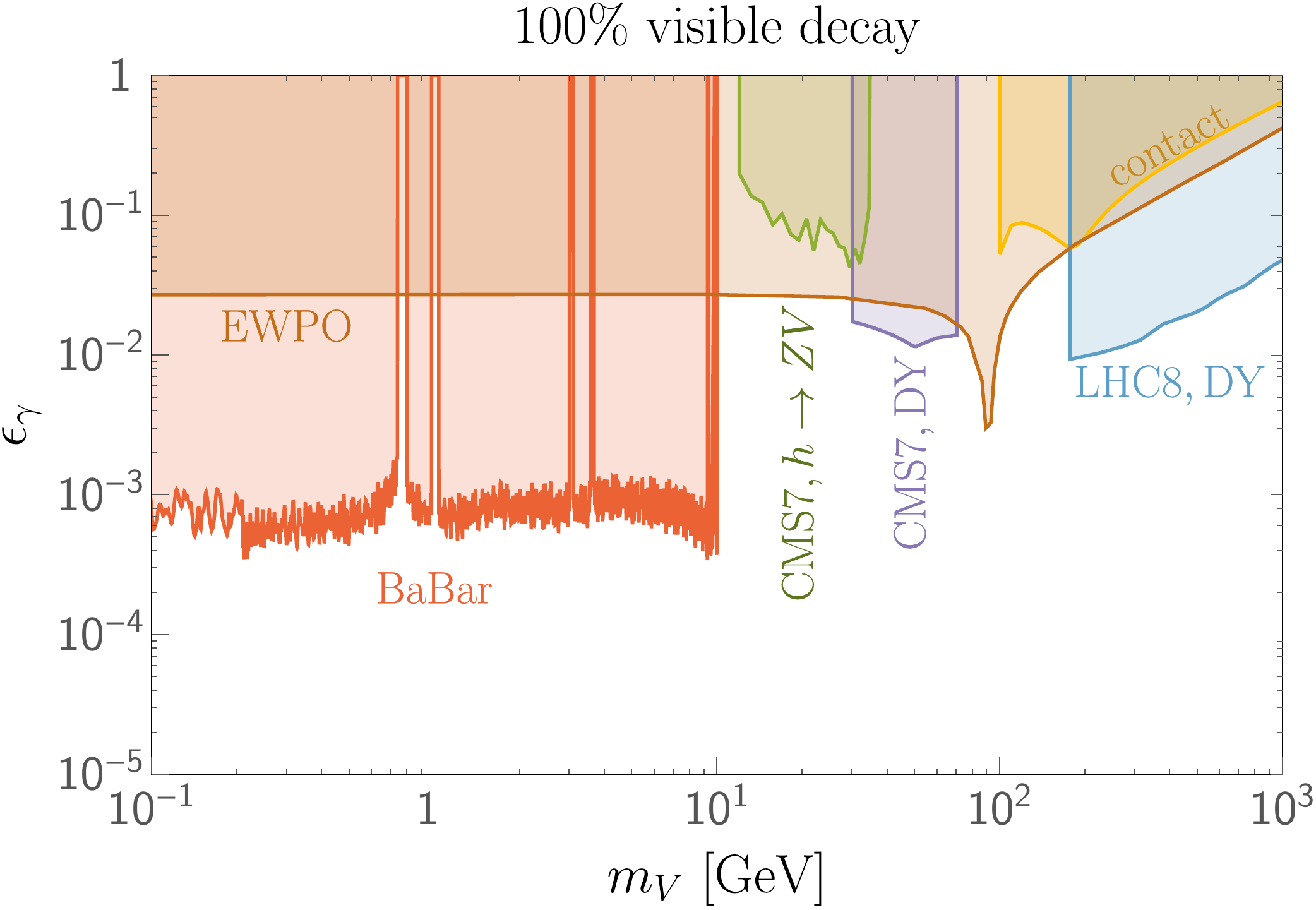}\hfill
   \includegraphics[clip,width=0.49\textwidth]{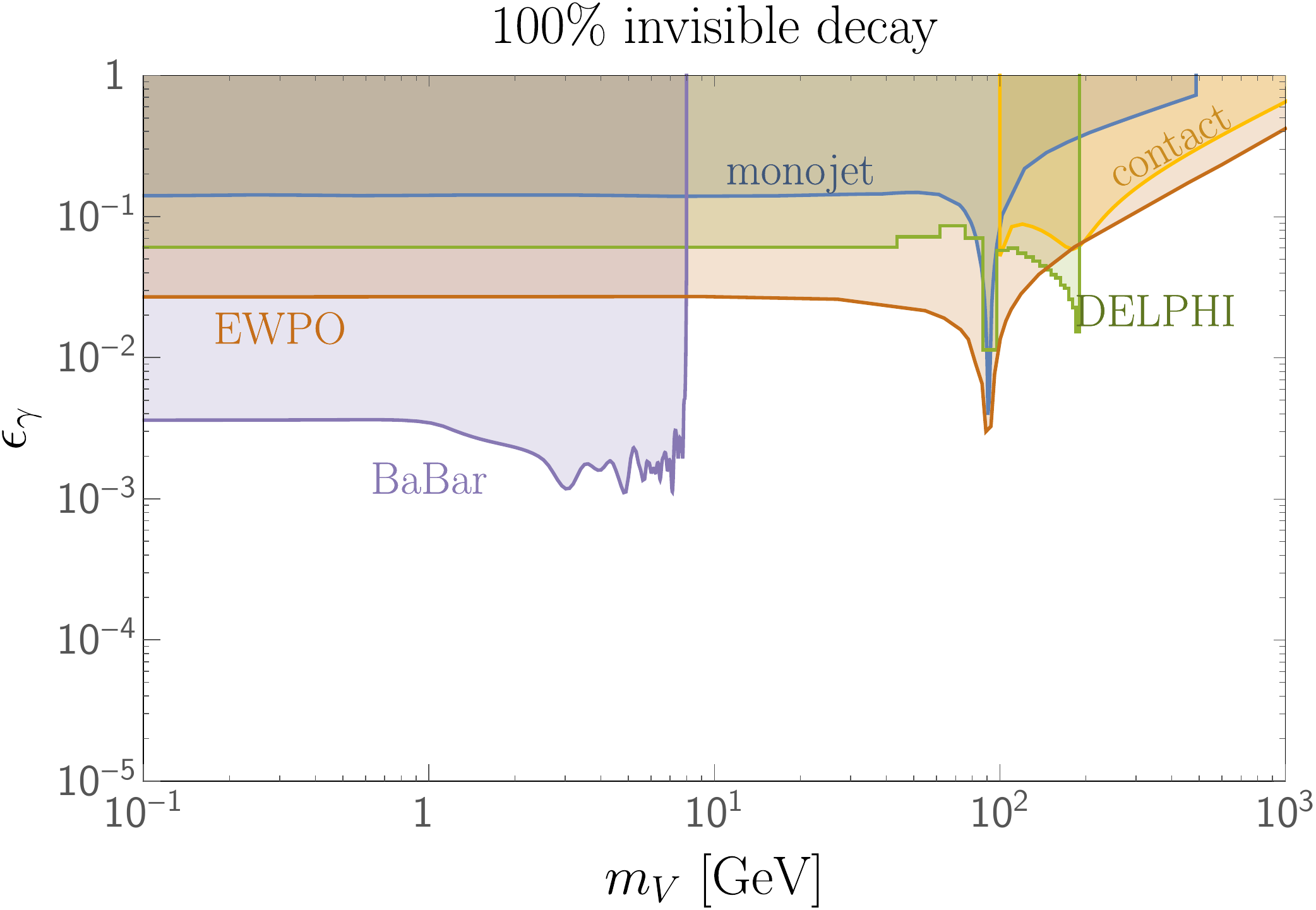}
  \caption{Constraints on $\epsilon_\gamma$ as a function of $m_V$ from dark photon decays; see text for details. {\bf Left:} 100\% visible decays (including decays to
  $\nu \bar \nu$); {\bf Right:} 100 \%  invisible decays.}
  \label{fig:DPlim}
\end{figure}

In order to understand the range of parameters viable for the SIMP mechanism, we now assemble the current experimental limits on dark photons.
There are two extreme cases of the dark photon decays to be considered: (I)~100\% branching fraction into Standard Model particles (left panel of
Fig.~\ref{fig:DPlim}), and (II)~100\% branching fraction into invisible particles (right panel of Fig.~\ref{fig:DPlim}).

Most of the limits presented in Fig.~\ref{fig:DPlim} are reproduced from Refs.~\cite{Izaguirre:2013uxa,Essig:2013vha,Curtin:2014cca}. However, a
few comments are in order.  First, we do not consider the case of a dark photon lighter than twice the dark pion.  The reason is that if the coupling of this process is strong enough to achieve kinetic equilibrium between the two sectors at the time of the freeze-out, it would cause the
annihilation rate of $\pi\pi\to \pi V$ to dominate over the $\pi\pi\pi\to \pi\pi$ process required for the SIMP mechanism~\cite{Lee:2015gsa}. The
constraint from electroweak precision observables (labeled {\tt EWPO}, brown)~\cite{Essig:2013vha} is independent of how the dark photon decays and is shown in both panels of Fig.~\ref{fig:DPlim}. Another constraint relevant to both visible and invisible decays comes from the search for a contact four-fermion operator at LEP II~\cite{Schael:2013ita} (labeled {\tt contact}, orange), which was not discussed in
Refs.~\cite{Izaguirre:2013uxa,Essig:2013vha,Curtin:2014cca}.  This process is nominally weaker than the limit from EWPO. We include it, however,
since it is expected to improve at the ILC by roughly an order of magnitude~\cite{Baer:2013cma} (see Fig.~\ref{fig:SIMPfuture}).

Bounds from visible decays\footnote{We define `visible decays' to include decays into neutrinos in this paper.} of a dark photon are shown in the left panel of Fig.~\ref{fig:DPlim}. Starting from the low $m_V$ region, the first
constraint (labeled {\tt BaBar}, red) is from a dedicated search for $e^+ e^- \rightarrow \gamma V$ followed by $V \rightarrow e^+ e^-, \mu^+ \mu^-$ at BaBar~\cite{Lees:2014xha}. The next constraint (labeled {\tt CMS7,$h\rightarrow ZV$}, green) is the reinterpretation of the study of $h \rightarrow Z^{(*)} Z^* \rightarrow 4\ell$ by CMS~\cite{CMS:xwa} as a search for $h \rightarrow Z^{(*)} V \rightarrow 4\ell$~\cite{Curtin:2014cca}. The third
bound (labeled {\tt CMS7,DY}, purple) is from the search for Drell--Yan production of lepton pairs performed by CMS in Ref.~\cite{Chatrchyan:2013tia}. The right most bound (labeled {\tt LHC8,DY}, blue) also comes from the search for Drell--Yan production of lepton pairs in Ref.~\cite{Cline:2014dwa},
using the ATLAS data~\cite{ATLAS:2013jma}.

Similarly, there are bounds specific for the invisible decays of the dark photon; these are shown in the right panel of Fig.~\ref{fig:DPlim}.
Starting from the low $m_V$ region, the first one (labeled {\tt BaBar}, purple) is from a search at BaBar for $\Upsilon(3S) \rightarrow \gamma A^0$,
where $A^0$ is a pseudo-scalar boson that decays invisibly~\cite{Aubert:2008as}.  Note that $\Upsilon(3S) \rightarrow \gamma V$ is forbidden by
charge conjugation invariance, but the limit can be re-interpreted as that on the continuum production $e^+ e^- \rightarrow \gamma V$ followed by
$V \rightarrow {\rm invisible}$~\cite{Izaguirre:2013uxa,Essig:2013vha}.  The next bound (labeled {\tt monojet}, blue) is for $pp \rightarrow {\rm
jet}+V$, $V\rightarrow \mbox{invisible}$.  Here the limit is scaled from the coupling to the up-quark in Ref.~\cite{Shoemaker:2011vi} that used
data from CDF and ATLAS (7~TeV).  This bound is currently subdominant to other constraints, but is expected to be improved with higher luminosity
and energy at the LHC. Finally, the limit from the mono-photon search at DELPHI~\cite{Abdallah:2008aa} (labeled {\tt DELPHI}, green) is reinterpreted, allowing for 2$\sigma$ fluctuations in each energy bin.

All limits are expected to be improved in the future by the LHC Run-2 and beyond, SuperKEKB, ILC and/or a 100~TeV $pp$ collider.  The limits on a
given model depend on the branching fractions of the dark photon into visible particles vs.~invisible modes.  These branching ratios will be
studied in Section~\ref{sec:decays}. Future prospects will be discussed in Section~\ref{sec:future}.

\section{The dark photon in SIMP Models}\label{sec:simpU1}

We now study how the dark photon couples to the SIMP sector.  We start by identifying the $U(1)_D$ subgroup of the unbroken global flavor symmetry
$H$, in the strongly coupled SIMP theory in Section~\ref{ssec:gaugeU1}. We then study two basic constraints for the SIMP setup in this
realization: (I)~kinetic equilibrium (thermalization) between the Standard Model and SIMPs during freeze-out (Section~\ref{ssec:theq}); and
(II)~dominance of the $3\rightarrow 2$ annihilation of SIMPs over the $2\to2$ annihilation into SM particles, $\pi\pi \rightarrow V^* \rightarrow
\ell^+ \ell^-$ (Section~\ref{ssec:ann}).

\subsection{Gauging $U(1)_D$}\label{ssec:gaugeU1}

Here we explore the embedding of the dark photon in the symmetry structure of the SIMP strongly coupled theories. The simplest possibility is to
gauge a $U(1)_D$ part of the unbroken flavor symmetry $H$ of the coset space $G/H$.  In principle one can also consider gauging a broken symmetry,
in which case one of the pions is eaten by the gauge boson.  We consider the case of unbroken generators here only for simplicity.
Ref.~\cite{Lee:2015gsa} embedded the $U(1)_D$ in an $SU(N_c)$ gauge theory with $N_f=3$; here we demonstrate the possibility of embedding the
$U(1)_D$ symmetry in general $Sp(N_c)$, $SU(N_c)$ and $SO(N_c)$ gauge theories with $N_f$ flavors.

Which subgroup of $H$ should be gauged?  There are two requirements behind the choice.  The first one is that the pions do not decay into dark photons. This can be guaranteed if all the pions transform non-trivially under part of the unbroken flavor symmetry.  The other condition is to maintain the near degeneracy of quarks, so that at least five different pions participate in the $3\rightarrow 2$ process via the WZW term.  The $U(1)_D$ coupling renormalizes the quark masses,
\begin{equation}
  \Delta m_q = \frac{3\alpha_D}{2\pi} Q_q^2 m_q \log \frac{M}{\Lambda}\,,
\end{equation}
where $\alpha_D\equiv e_D^2/(4\pi)$ is the dark coupling strength, $M$ is a UV cutoff such as the Planck scale, $\Lambda$ is the confining scale, and $Q_q$ is the $U(1)_D$-charge of quark $q$.  If $Q_q^2$ is not common to all quarks, quark masses are split at the $\sim15\%$ level for $\alpha_D \approx \alpha$ (with $\alpha$ the electroweak coupling strength), which may or may not be acceptable.  For this reason, and for simplicity, we choose $U(1)_D$ so that all quarks have the same charge, up to a sign.

We first consider $Sp(N_c)$ gauge theories, and the resulting flavor coset
 $SU(2N_f)/Sp(2N_f)$.  The pions transform in the anti-symmetric tensor
of $Sp(2N_f)$.   This agrees with the intuition that the pions are the
$S$-wave spin-zero bound state of massive Majorana quarks, with gauge indices contracted by the anti-symmetric symplectic tensor, requiring the
anti-symmetric combination of flavor indices. If the quarks are degenerate, the quark mass term also respects $Sp(2N_f)$, and hence the degeneracy is preserved by the flavor
symmetry.  The $ U(1)_D$ is embedded in the global symmetry as
\beq
 SU(N_f) \times U(1)_D \subset Sp(2N_f)
\eeq
The quarks transform  as $(\Yfund, +1)$ and $(\overline{\Yfund}, -1)$ multiplets under $SU(N_f) \times U(1)_D $.  The pions decompose into
representations $A(\Yasymm\, , +2)$, $A^*(\overline{\Yasymm}\, ,-2)$, and $\Omega({\rm adj}, 0)$,
\begin{equation}
  \pi = \left(
    \begin{array}{cc}
      \Omega & A\\
      -A^* & \Omega^T
    \end{array}
  \right)\,,
\end{equation}
and since the $SU(N_f)$ symmetry is exact, neither the anti-symmetric tensor $A$ nor the adjoint $\Omega$ can decay, and all pions are stable.

When the $U(1)_D$ symmetry is gauged, the WZW-action includes additional gauge-interactions. Up to linear terms in the $U(1)_D$ gauge boson $\cal
A$, the gauged WZW-action is
\beq\label{WZW1}
\Gamma_{\rm WZW}^{(1)} &=& \frac{-i 5 N_c }{240 \pi^2}   \int_{M^4} e_D {\cal A} \Tr\, Q \left[(d\Sigma \Sigma^{-1})^3 - (\Sigma^{-1}d\Sigma )^3
\right]  \nonumber \\
  &=& \frac{-i 2 e_D N_c }{3 \pi^2 f_\pi^3}  \int d^4x \epsilon^{\mu\nu\rho\sigma}  {\cal A} _\mu \tr[ \partial_\nu  \Omega \partial_\rho \Omega
  \partial_\sigma \Omega  + \partial_\nu  \Omega \partial_\rho A   \partial_\sigma A^\dagger ]\,.
\eeq
Here, `Tr' in the first line is the trace over $2N_f \times 2N_f$ matrices, while `tr' in the second line is the trace over $N_f \times N_f$ matrices.
Semi-annihilation channels of $\pi \pi \rightarrow \pi V$ are then present, and may spoil the SIMP mechanism. We suppress these processes by
taking $m_V \gsim 2m_\pi$, rendering the process kinematically inaccessible during freeze-out (including effects of the thermal tail). Similarly, semi-annihilation interactions will be
present from the gauged WZW action in the other gauge theories, and so we will always take $m_V\gsim 2m_\pi$.

\begin{table}[t]
\begin{center}
\begin{tabular}{ |c||c|c|}
\hline\hline \rule{0pt}{1.2em}
 ~	&  $U(1)_D\subset H$ embedding 		& $\pi$ representations\\ \hline\hline
$Sp(N_c)$			&  $\begin{array}{c}SU(N_f)\times U(1)_D \\ \subset Sp(2 N_f)\end{array}$ 	& $(\Yasymm\, , +2)\oplus(\overline{\Yasymm}\,
,-2)\oplus({\rm adj}, 0)$ \\ \hline
$SO(N_c)$			&  $\begin{array}{c}SU(N_f/2)\times U(1)_D\\ \subset SO(N_f)\end{array}$ & $(\Ysymm, +2) \oplus (\overline{\Ysymm}, -2) \oplus
({\rm adj}, 0)$ \\ \hline
$SU(N_c)$			&  $\begin{array}{c}SU(N_1)\times SU(N_2)\times U(1)_D\\ \subset SU(N_f)\end{array}$ & $\begin{array}{c}(\Yfund,
\overline{\Yfund}, 2) \oplus (\overline{\Yfund}, \Yfund, -2) \oplus \\  ({\rm adj}, 1, 0) \oplus (1, {\rm adj}, 0) \oplus (1, 1, 0)\end{array}$ \\
\hline
\hline
\end{tabular}
\caption{Gauging the $U(1)_D$ subgroup of the unbroken flavor symmetry, $H$, for the $Sp(N_c)$, $SO(N_c)$ and $SU(N_c)$ gauge theories. }
\label{tab:gauging}
\end{center}
\end{table}

For the $SO(N_c)$ gauge theory with the $SU(N_f)/SO(N_f)$ coset, if the number of flavors is even $N_f = 2 n_f$, the gauged $U(1)_D$ is embedded
as
\beq
SU(n_f) \times U(1)_D \subset SO(N_f), ~~~~~~ n_f = N_f/2
\eeq
The quarks transform as $(\Yfund, +1) \oplus (\overline{\Yfund}, -1)$ under $SU(n_f)\times U(1)_D$, and the pions as $(\Ysymm, +2) \oplus (\overline{\Ysymm},
-2) \oplus ({\rm adj}, 0)$. This set of quantum numbers is consistent with the $S$-wave spin-zero bound states of massive Majorana quarks, with
gauge indices contracted by the symmetric Kronecker-delta tensor, requiring the symmetric combination of flavor indices.

The generalization of the $U(1)_D$ gauging in the $SU(N_c)$ gauge theory in Ref.~\cite{Lee:2015gsa} is the following.  The $U(1)_D$ is embedded in
the unbroken group as
\beq
 SU(N_1) \times SU(N_2) \times U(1)_D \times U(1)_B  \subset SU(N_f) \times U(1)_B
 \eeq
where the $U(1)_D$ generator is
\begin{equation}\label{eq:Qsun}
  Q={\rm diag}(\underbrace{+1, \cdots, +1}_{N_1}, \underbrace{-1, \cdots, -1}_{N_2})
\end{equation}
with $N_1+N_2=N_f$.  The pion states transform as $(\Yfund, \overline{\Yfund}, 2) \oplus (\overline{\Yfund}, \Yfund, -2) \oplus ({\rm adj}, 1, 0) \oplus (1, {\rm adj}, 0) \oplus (1, 1, 0)$ under the $SU(N_1)\times SU(N_2)\times U(1)_D$.

The pions transforming in the non-trivial representations of the unbroken flavor group are protected from decay. The singlet pion in the $SU(N_c)$
gauge theory, however, is not protected, but must be stable at the time of freeze-out; otherwise the other pions can scatter into this neutral
state which then decays, eventually depleting the pion abundance.  This issue cannot be avoided even if the singlet pion decays after freeze-out,
because then the pion will decay after BBN and dissociate light elements.  Therefore, the pion must be stable on cosmological time scales.  The
above choice of $U(1)_D$, Eq.~\eqref{eq:Qsun}, avoids the possibility of the singlet neutral pion $\pi^0$ decaying into $\pi^0 \rightarrow V^*V^*
\rightarrow e^+ e^- e^+ e^-$ via the axial anomaly.  This is because the singlet pion corresponds to the generator
\begin{equation}
  T^0 = \sqrt{\frac{2}{(N_1+N_2)N_1N_2}}\ (\underbrace{+N_2, \cdots,
  +N_2}_{N_1}, \underbrace{-N_1, \cdots, -N_1}_{N_2})\,,
\end{equation}
and the triangle anomaly diagram vanishes, ${\rm Tr} T^0 Q Q=0$.  The Adler--Bardeen theorem guarantees that this cancellation persists to all
orders in perturbation theory~\cite{Adler:1969er}.  Inserting more photons does not make $\pi^0$ unstable.  This is because charge conjugation
invariance requires an even number of photons, while ${\rm Tr}\, T^0 Q^{2n}=0$ for any $n \in {\mathbb Z}$. Double traces ${\rm Tr}\, T^0 Q^{2n-1}
{\rm Tr}\, Q^{2m-1}$ need not vanish, but the corresponding Feynman diagram, that may generate the effective operator, vanishes because of charged
conjugation invariance of each fermion loop (Furry's theorem).
This is an intriguing case where a particle with no conserved quantum number appears to be stable. We note that the singlet $\pi^0$ may decay through a higher dimension operator in the QCD-like theory, induced by some heavy particles, leading to a
signal of late-time decay in the galactic halo.

The non-degeneracy among the charged and neutral pions, generated below the confinement scale, is of the order of $\Delta m_\pi^2 \approx e_D^2
f_\pi^2$. Sufficient degeneracy between the neutral and charged pions should be maintained in order to achieve the $3\to2$ annihilation through
the WZW term.

Having established the embedding of the dark photon in the confining SIMP theories, we now proceed to address the conditions of the SIMP mechanism
in the presence of the dark photon.

\subsection{SIMP conditions: Thermalization}\label{ssec:theq}

A crucial requirement of the SIMP mechanism is that the entropy is dumped out of the SIMP sector while the SIMPs freeze-out through the
$3\rightarrow 2$ process.  We demand that the scattering rate between the SIMPs and the light Standard Model leptons ($e^\pm$, $\nu$, $\bar{\nu}$)
is faster than the expansion rate at the freeze-out temperature $T_f$, described by $x_f = m_\pi / T_f \approx 20$.

\begin{figure}
  \centering
  \includegraphics[clip,width=0.8\textwidth]{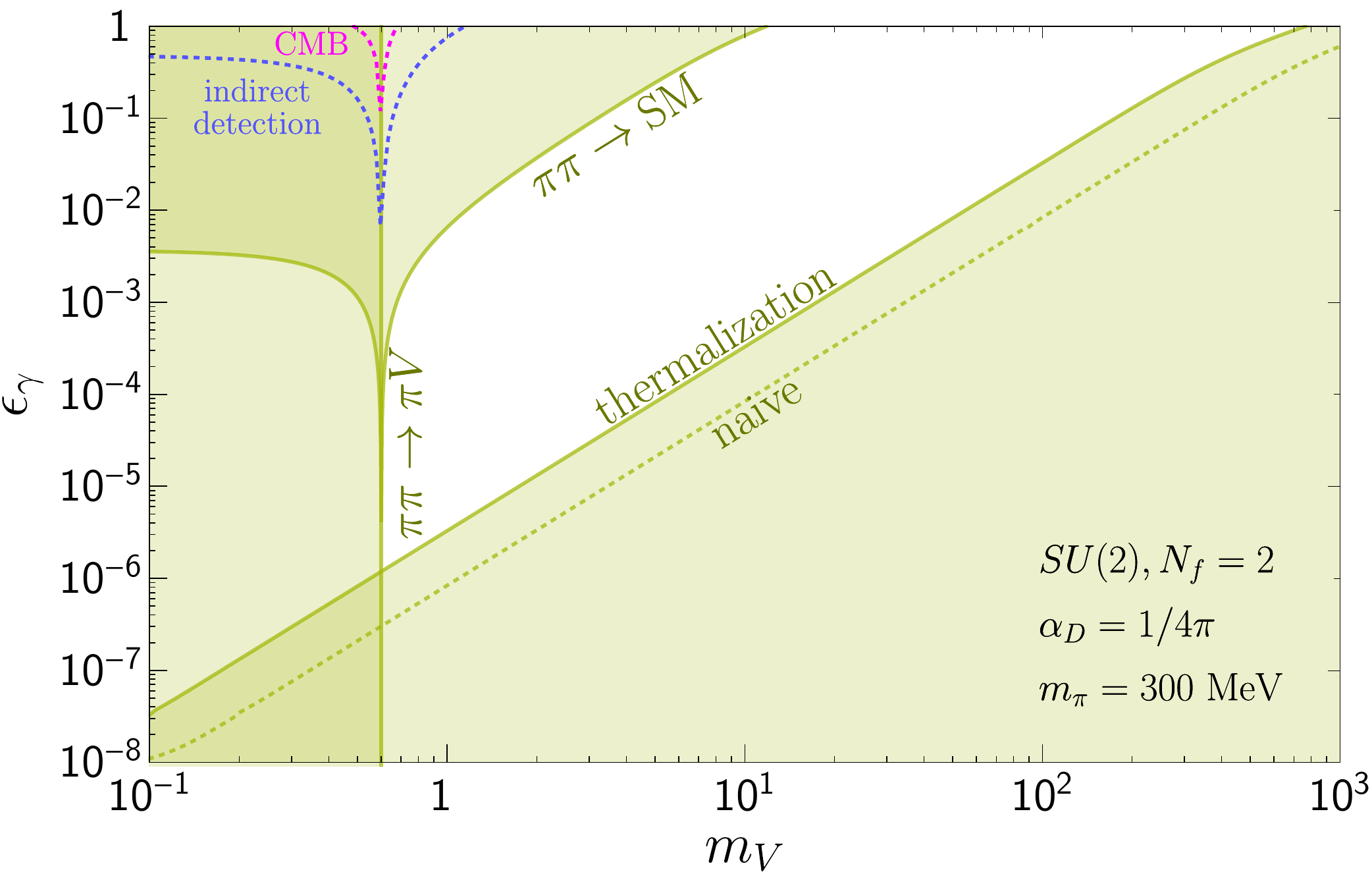}
  \caption{Constraints on the dark photon parameter space from requiring:
  (I)~kinetic equilibrium between the Standard Model and SIMPs during freeze-out (labeled {\tt thermalization}, and compared to the {\tt naive} thermalization estimate);
    (II)~dominance of the $3\rightarrow 2$ annihilation of SIMPs over
    $\pi\pi \rightarrow V^* \rightarrow \ell^+ \ell^-$ (labeled
    $\pi\pi\rightarrow$SM), and (III)~subdominance of the
    semi-annihilation process $\pi \pi \rightarrow \pi V$ (labeled as
    such).  We used the SIMPlest model of an $SU(2) = Sp(2)$ gauge theory
    with $N_f = 2$, $\alpha_D = 1/4\pi$, and $m_\pi = 300$~MeV. (Suppressed) constraints from CMB data and indirect detection are shown in the
    dashed magenta and dashed blue curves, respectively.  }
  \label{fig:Thermalization}
\end{figure}

The scattering cross section $\pi \ell \rightarrow \pi \ell$, where $\ell$ is a SM lepton (assumed massless), is
\begin{eqnarray}\label{eq:sigann}
  \sigma^{f}_{\rm scatt} v_{rel}
  &=& 8\pi \alpha_D \alpha \frac{\sum Q_\pi^2}{N_\pi}
  \frac{1}{c_\chi^2}
  \left[ - \left( \frac{c_\zeta^2}{m_V^2} + \frac{s_\zeta^2}{m_Z^2} \right)
  c_W t_\chi Q_f \right. \nonumber \\
  & & \left. +
  \left(\frac{c_\zeta (s_\zeta +s_W t_\chi c_\zeta)}{m_V^2}
  - \frac{s_\zeta (c_\zeta - s_W t_\chi s_\zeta)}{m_Z^2}  \right)
  \frac{1}{s_W c_W} (I_3^f - Q_f s_W^2) \right]^2 E_\ell^2
\end{eqnarray}
at threshold.  We disagree with Eq.~(18) in Ref.~\cite{Lee:2015gsa} where the mixing with the photon and $Z$ are added incoherently in the cross
section, or no $Z$-boson exchange diagram is considered.  The above expression Eq.~\eqref{eq:sigann} should be summed over $f= e_R^\pm$,
$e_L^\pm$, $\nu_{e,\mu,\tau},\bar{\nu}_{e,\mu,\tau}$, where
$\langle E_\ell^2 n_\ell \rangle = \frac{12 \zeta(5)}{\pi^2} \frac{15}{16} T_f^5$. A naive estimate on the overall scattering rate can be obtained by requiring it  be larger than the expansion
rate at freezeout,
\begin{equation}
  \Gamma_{\rm scatt} =
   \sum_f
    \frac{\sigma^f_{\rm scatt} v_{rel}}{E_f^2} \frac{12
      \zeta(5)}{\pi^2} \frac{15}{16} T_f^5 \quad \lesssim\quad H(T_f)\,.
\end{equation}
The resulting constraint is depicted by the dotted line of Fig.~\ref{fig:Thermalization} (labeled {\tt naive}).

A more detailed study of the rate of energy
transfer between the SM and SIMP sectors results in a slightly more stringent constraint, strengthening the depicted curve~\cite{Kuflik:2015isi}.
The rate at which energy is lost for a dark matter particle scattering with a bath fermion via the process $\pi_1 f_1 \to \pi_2 f_2$ is
\begin{equation} \label{kineqint}
\dot{E}= \frac{1}{2 E_{\pi_1}}\int  d\Pi_{f_1} d\Pi_{\pi_2} d\Pi_{f_2} (2\pi )^4\delta ^4\left(p_{\pi_1}+p_{f_1}-p_{\pi_2}-p_{f_2} \right) f_{f_1}(E_{f_1}) \left(  E_{\pi_2} -E_{\pi_1} \right) \left|{\cal M}\right|^2\,,
\end{equation}
where $ d\Pi_i =g_i d^3 p_i /[(2\pi)^32 E_i]$ and $\left|{\cal M}\right|^2$ is averaged over initial and final state degrees of freedom. This should be compared to the rate at which kinetic-energy per-particle is being changed by the expansion of the universe, while chemical equilibrium is maintained by the $3\to 2$ self-annihilations.  The rate at which the mass of the disappearing dark matter is transferred to kinetic energy in the dark-matter bath, by the $3\to 2$ process is,
\beq
\dot{E} &=& \frac{1}{n_\pi} \frac{\partial (m_\pi n_\pi) }{\partial t} \simeq - \frac{H m_\pi^2}{T}\,.
\eeq
Therefore, in order for kinetic equilibrium to be maintained during freeze-out, the overall scattering rate must satisfy
\begin{equation}
  \frac{5\zeta(5)}{4} m_\pi  \Gamma_{\rm scatt}  \lesssim \frac{H(T_f)  m_\pi^2}{T_f}\,,
\end{equation}
where the coefficient of the left-hand-side is determined by performing the integral in Eq.~(\ref{kineqint}). The resulting constraint is depicted in the bottom shaded region of Fig.~\ref{fig:Thermalization}, stronger by a factor of $\sim 3$ compared to the naive estimate.

\subsection{SIMP conditions: Annihilation}\label{ssec:ann}

The second condition for SIMP dark matter is that at the time of freeze out, no new SIMP annihilations beyond the $3\to2$ process should be
dominantly active.  To this end, we require the $2\to2$ annihilation of $\pi \pi \rightarrow V^* \rightarrow e^+ e^-$ to be out of equilibrium at
freeze-out.  The annihilation cross section is given by
\begin{eqnarray}
  \lefteqn{
  \sigma_{\rm ann}^\ell =
  \beta \frac{8\pi\alpha_D\alpha}{3 c_\chi^2}  \frac{1}{N_\pi^2}\sum_\pi Q_\pi^2 m_\pi^2
  \left[
  - \left( \frac{c_\zeta^2}{m_V^2-s} + \frac{s_\zeta^2}{m_Z^2-s} \right)
  c_W t_\chi Q_\ell \right. } \nonumber \\
& & \left. +
    \left(\frac{c_\zeta (s_\zeta +s_W t_\chi c_\zeta)}{m_V^2-s}
    - \frac{s_\zeta (c_\zeta - s_W t_\chi s_\zeta)}{m_Z^2-s}  \right)
    \frac{1}{s_W c_W} (I_3^f - Q_\ell s_W^2) \right]^2\, ,
\end{eqnarray}
and the requirement that annihilations are off during freezeout is
\begin{eqnarray}
  \Gamma_{\rm ann}
  = \sum_{f=e^-_R, e^-_L, \nu_{e,\mu,\tau}}
      \frac{\sigma^\ell_{\rm ann}}{\beta} 3 x_f^{-1} Y_\pi s_f \quad \lesssim\quad H(T_f)\,.
\end{eqnarray}
Here,
\beq
  s_f = \frac{2\pi^2}{45} g_{*s} T_f^3\,,\quad {\rm and}\quad
  Y_\pi = \frac{n_\pi}{s} = 4.39 \times 10^{-10} \frac{\rm GeV}{m_\pi}\,,
\eeq
represent the entropy at freeze-out and the measured dark matter abundance, respectively.  The resulting constraint is depicted in the top shaded
region of Fig.~\ref{fig:Thermalization}.

The semi-annihilation process of $\pi \pi \rightarrow \pi V$ must also be subdominant to the $3\to2$ process during freeze-out. Since the
thermalization requirement bounds $\alpha_D \epsilon_\gamma^2$ from below, and the semi-annihilation process is proportional to $\alpha_D$, the
only way to suppress the semi-annihilation process while maintaining thermal equilibrium with the SM is to require a kinematic suppression,
$Y_\pi Y_V\lesssim Y_\pi^3$ with the yields $Y_{\pi,V} \sim e^{-m_{\pi,V}/T_f}$.  Therefore, we need $m_V \gtrsim 2 m_\pi$, compensating for the thermal tail.  This constraint is shown in the shaded region towards the left-hand side of Fig.~\ref{fig:Thermalization}.

In principle, stringent limits on light dark matter annihilations in the late Universe can exist~\cite{Hochberg:2014dra}. These come from
measurements of the~CMB power spectrum~\cite{Madhavacheril:2013cna} as well as from measurements of diffuse gamma-rays in the
galactic halo~\cite{Essig:2013goa}. In our case, the annihilation of dark pions through the $s$-channel exchange of a dark photon, $\pi^+ \pi^-
\rightarrow V^* \rightarrow e^+ e^-$, proceeds strictly in the $p$-wave, and the corresponding annihilation cross sections at later times are
suppressed by $v^2$. Consequently, the cross section is much smaller at the time of recombination and in the galactic halo today. The process
$\pi^+ \pi^- \rightarrow V^* V^* \rightarrow e^+ e^- e^+ e^-$ is also suppressed, by $\epsilon_\gamma^4 (\frac{\alpha}{2\pi}
\frac{m_\pi^2}{m_V^2})^2$. The strength of the CMB and indirect-detection data is shown in the dashed magenta and blue and curves in
Fig.~\ref{fig:Thermalization}, respectively. As is evident, these constraints play no role in the SIMP/dark-photon setup. Likewise, bounds from stellar cooling
are evaded for SIMP pion masses of order a few hundred MeV, as they (and the heavier mediator) are too heavy to be produced in white dwarfs or
supernovae. We learn that SIMP dark pions together with a dark photon mediator provides a light dark matter candidate that easily evades
astrophysical and cosmological constraints.

\section{Dark photon decays}\label{sec:decays}

The limits on the dark photon parameters discussed in Section~\ref{sec:darkphoton} were obtained for 100\% branching fraction either into visible or invisible modes.  In a given model, the branching fractions into each mode must be computed in order to properly identify the viable parameter space. This is a complicated, yet fascinating task, due to the strong dynamics of both the QCD and SIMP sectors.  We attempt the computation of partial widths in this section.

\subsection{Invisible width of dark photon}\label{ssec:inv}

The dark photon can decay into the dark sector.  In $e^+ e^-$ annihilations, the virtual dark-photon can create hadronic
states and probe the structure of the strongly coupled theory, determining properties such as  the number of colors, flavors and the electric charges of the states.  Given that the dark sector is a QCD-like strongly coupled gauge theory, the dark photon can produce vector resonances and hence perform spectroscopy of the dark sector, in a very similar way to spectroscopy of QCD resonances from virtual photons.  To the best of our knowledge, this important point has not been discussed in the dark photon literature.

It is an interesting theoretical challenge how to model the dark spectroscopy.  In what follows, we use a parametrization inspired by `soft wall' holographic QCD~\cite{Karch:2006pv} which mimics the resonance spectrum of radial excitations consistent with the Regge behavior of $m^2_n \propto (n+J)$. This tool, used to model strongly coupled gauge theories, will enable us to understand what region of parameter space is available, as well as to demonstrate in a concrete way the various aspects of the proposed SIMP spectroscopy.

\subsubsection{Parton-level widths}\label{sssec:parton}

When the dark photon mass is much higher than the dynamical scale of the hidden strong sector, the invisible decays of the hidden photon can be computed at parton-level into hidden quarks.  The width of the hidden photon to quark-pairs is
\begin{equation}\label{eq:gammainv}
  \Gamma(V \rightarrow q \bar{q}) = \left\{ \begin{array}{lcc}
  N_c N_f \Gamma_0 & & SU(N_c)\ {\rm or}\ Sp(N_c)\,,\\
  \frac{1}{2} N_c N_f \Gamma_0 & & SO(N_c)\,,
  \end{array} \right.
\end{equation}
where
\begin{equation}\label{eq:gamma0}
  \Gamma_0 = \frac{e_D^2}{12\pi} m_V = \frac{\alpha_D}{3} m_V\,,
\end{equation}
is the  two-body decay width into a single Dirac fermion.
We note that the analysis of Ref.~\cite{Lee:2015gsa} considered vector decays into pions, rather than into quarks.
In the $SU(N_c=3)$, $N_f=3$  model considered there, there are two charged scalars of charge two, and hence the invisible width to the charged pions is
$
  \Gamma(V \rightarrow \pi\pi) = 2 \Gamma_0 ,
$
while the invisible width into dark quarks is, according to Eq.~\eqref{eq:gammainv}, a factor of $3N_c/2$ larger. Of course, the parton-level calculation breaks down near the dynamical scale where the dark quarks form dark hadrons, and will be treated in the next section.

Despite the factor of $N_c N_f$ appearing in Eq.~\eqref{eq:gammainv}, we do not expect the invisible width to be very large.  For instance, requiring that the $U(1)_D$ coupling does not hit a Landau pole below the Planck scale gives
\begin{equation}\label{eq:pert}
  e_D^2 < \frac{8\pi^2}{b_0 \log (M_{\rm Pl}/\Lambda)}\ ,
\end{equation}
where
\begin{equation}\label{eq:b0}
  b_0 = \left\{ \begin{array}{lcc}
                  \frac{4}{3} N_c N_f & & SU(N_c)\mbox{ or }Sp(N_c)\,,\\
                  \frac{2}{3} N_c N_f & & SO(N_c)\,.
                \end{array} \right.
\end{equation}
Therefore,
\begin{equation}
  \Gamma(V \rightarrow q\bar{q}) < \frac{\pi}{2\log
    (M_{\rm Pl}/\Lambda)}\ m_V \simeq 0.037 m_V\,.
\end{equation}
The perturbativity constraint on the $U(1)_D$ coupling, Eq.~\eqref{eq:pert}, also ensures that the dark pions are sufficiently degenerate for the $3\to2$ (co-annihilation) process to proceed.


\subsubsection{Vector resonances}\label{ssec:resonances}

As the mass of the hidden photon approaches the confinement scale, the
effects of hadronic vector resonances of the strongly coupled theory
on the invisible decay width become important. However, we cannot compute the strongly coupled physics from first principles.
For the purpose of calculating the invisible branching ratios of the
hidden photon, and illustrating spectroscopy at future $e^+ e^-$
colliders, the resonance structure needs to be modeled in a reasonable
way.  For the discussions here, we focus on $Sp(N_c)$ gauge theories,
while $SO(N_c)$ and $SU(N_c)$ gauge theories are discussed in
Appendix~\ref{sssec:holo}.

The $\rho$ mesons are $L=0$, $S=1$ bound
states of the Majorana quarks whose gauge indices are contracted by
the (anti-symmetric) symplectic tensor.  Therefore the flavor indices
transform as a symmetric tensor of the unbroken $Sp(2N_f)$.  Under
$SU(N_f)\times U(1)_D \subset Sp(2N_f)$, they decompose as
$(\Ysymm\, , +2)\oplus(\overline{\Ysymm}\, ,-2)\oplus({\rm adj},
0)\oplus({\rm 1},0)$.  The dark photon couples to the singlet representation $\rho({\rm 1},0)$.

The width of the dark photon can be obtained from the vector-vector
correlation function or vacuum polarization diagram,
\begin{equation}
  m_V \Gamma_V = N_f e_D^2 m_V^2 {\Im m} \Pi(m_V^2)\,,
\end{equation}
where the factor of $N_f$ comes from the normalization of the $U(1)_D$
charge, and the vacuum polarization function $\Pi(q^2)$ is defined by
\begin{equation}
  \int d^4x e^{i q\cdot x} \langle J_\mu^a(x) J_\nu^b(0) \rangle
  = i \delta^{ab} (q_\mu q_\nu - g_{\mu\nu}) \Pi(q^2).
\end{equation}

In the strongly coupled regime and large $N_c$, the same vacuum
polarization function should be obtained by summing over the vector
resonances \cite{'tHooft:1973jz}.  The vector resonances $\rho_n$ are
characterized by their masses and decay constants with the
corresponding vector currents $J_\mu^a$,
\begin{equation}
  \langle 0 | J_\mu^a | \rho_n^b \rangle
  = \delta^{ab} F_{\rho_n} \epsilon_\mu\,.
\end{equation}
Casting this into an effective Lagrangian  using its
source gauge field $J_\mu^a {\cal A}^{\mu a}$,
\begin{equation}
  {\cal L}_{\it eff}
  = -\frac{1}{4 e_D^2} {\cal A}_{\mu\nu}^a {\cal A}^{\mu\nu a}
  + \sum_n -\frac{1}{4} {\cal \rho}_{\mu\nu}^a {\cal \rho}^{\mu\nu a}
  + \frac{F_{\rho_n}}{2m_{\rho_n}^2} {\cal A}_{\mu\nu}^a \rho_n^{\mu\nu a}
  + \frac{1}{2} m_{\rho_n}^2 \rho_{n\mu}^a \rho_n^{\mu a}\ .
\end{equation}
The vacuum polarization diagram is obtained by integrating out
$\rho_n$ and differentiating the resulting action with respect to
${\cal A}_\mu^a$ twice (right-most diagram in Fig.~\ref{ADSQCD} in Appendix~\ref{sssec:holo}),
\begin{eqnarray}
  (q_\mu q_\nu - g_{\mu\nu}) \Pi(q^2)
  &=&  \sum_n \left( \frac{F_{\rho_n}}{m_{\rho_n}^2} \right)^2
      (q_\mu q_\kappa - g_{\mu\kappa}q^2)
      \frac{-g^{\kappa\lambda}+q^\kappa q^\lambda/m_{\rho_n}^2}{q^2-m_{\rho_n}^2+i\Gamma_n^{\rho} m_{\rho_n}}
      (q_\lambda q_\nu - g_{\lambda\nu}q^2) \nonumber \\
  &=&  -\sum_n \left( \frac{F_{\rho_n}}{m_{\rho_n}^2} \right)^2
      (q_\mu q_\kappa - g_{\mu\kappa}q^2)
      \frac{q^2}{q^2-m_{\rho_n}^2++i\Gamma_n^{\rho} m_{\rho_n}}\ ,
\end{eqnarray}
leading to
\begin{equation}\label{eq:Pi}
  \Pi(q^2) = -\sum_n \frac{F_{\rho_n}^2}{m_{\rho_n}^2}
  \frac{q^2}{(q^2-m_{\rho_n}^2++i\Gamma_n^{\rho} m_{\rho_n})m_{\rho_n}^2}\ .
\end{equation}

In order to perform the sum, the spectrum and decay constants of the $\rho$-mesons are needed. To calculate these, we borrow results from the `soft-wall' holographic QCD~\cite{Karch:2006pv}, the minimal details of which are reviewed in
Appendix~\ref{sssec:holo}, and state the main results here.  Holographic QCD identifies $\rho$-mesons as the gauge bosons of the gauge flavor symmetry in the AdS bulk, which precisely agrees with the quantum numbers discussed above.  The spectrum and decay constants are given by
Eqs.~\eqref{eq:mn} and~\eqref{eq:Fn} (in units of AdS curvature),
\begin{eqnarray}
  m_{\rho_n}^2 = 4n\,,~~~~~~~~~~ \frac{F_{\rho_n}^2}{m_{\rho_n}^2} = \frac{2}{g_5^2}\,,~~~~~~~~~~ g_5^2 = \frac{12\pi^2}{N_c}\,,~~~~~~~~~~ n=1,2,...
\end{eqnarray}
Performing the sum over the tower of $\rho$-meson resonances in Eq.~(\ref{eq:Pi}), gives
\begin{eqnarray}
  \Pi(q^2)-\Pi(0)
  &=& - \frac{N_c}{24\pi^2}
      \left( 1- \frac{1}{\pi} \tan^{-1} (N_f-1)
      \frac{g_5^2 }{96\pi} \right)^{-1}
      H_{-q_{\it eff}^2/4}\, ,
\end{eqnarray}
with $H$ the harmonic number [see Eq.~\eqref{eq:Har}] and
\begin{equation}\label{eq:qeff}
  q_{\it eff}^2 = q^2 \left(1 + i \frac{\Gamma_n^\rho}{m_{\rho_n}}
    (m_\rho^2 = q^2) \right)\,.
\end{equation}
With simplifying assumptions given in Eq.~\eqref{eq:Gammarhon},
the $\rho \to \pi \pi $ partial widths are given by
\begin{equation}
  \Gamma_n^\rho =(N_f-1) \frac{\beta^3}{96\pi} g_5^2 m_{\rho_n}\,,
  \qquad \beta=\sqrt{1-\frac{4m_\pi^2}{m_{\rho_n}^2}}\, .
\end{equation}

\subsection{Visible widths of dark photon}\label{ssec:vis}

The decay of the dark photon into the visible sector can be computed
reliably (see {\it e.g.} Ref.~\cite{Batell:2009yf}).  For the leptonic final states:
\begin{equation}
  \Gamma(V \rightarrow \ell^+ \ell^-)
   = \frac{\alpha \epsilon_\gamma^2}{\alpha_D}
   \frac{  \beta(3-\beta^2)}{2}  \Gamma_0 \, ,~~~~~~  \beta = \sqrt{1 - \frac{4m_\ell^2}{m_V^2} }
\end{equation}
where $\Gamma_0 \equiv {\alpha_D m_V}/{3}$ as in Eq.~\eqref{eq:gamma0}. The hadronic width is modulated by the resonance structure of QCD.  Using the standard notation of the hadronic cross section in $e^+ e^-$ annihilations at center of mass energy $\sqrt{s}$,
\begin{equation}\label{eq:Rs}
R(s) \equiv \frac{\sigma(e^+ e^- \rightarrow
\mbox{hadron})}{\sigma(e^+ e^- \rightarrow \mu^+ \mu^-)_0}\,,
\end{equation}
where the subscript $0$ refers to the lowest order QED calculation for massless muons,
\beq
\sigma(e^+ e^- \rightarrow \mu^+ \mu^-)_0 = \frac{4\pi\alpha^2}{3s}\,,
\eeq
the hadronic width is
\begin{equation}
  \Gamma(V \rightarrow \mbox{hadrons})
  =  \frac{\alpha \epsilon_\gamma^2}{\alpha_D}
    R(m_V^2)\Gamma_0 \,.
\end{equation}
We use data of $R(s)$ from Ref.~\cite{Agashe:2014kda}.

When $m_V$ is much higher than the QCD scale and is comparable to or above $m_Z$, the full expression of Eq.~(\ref{eq:LD}) should be used for each Weyl fermion in the Standard Model,
\begin{eqnarray}\label{eq:Vff}
  \Gamma_{V\to f \bar{f}}
  &=& \frac{\alpha}{\alpha_D} \frac{1}{2} \left[
      Q_f \epsilon_\gamma + \frac{1}{s_W c_W}
      (I_3^f-Q_fs_W^2) \epsilon_Z \right]^2
         \frac{  \beta(3-\beta^2)}{2}  \Gamma_0 \, .
\end{eqnarray}
For quark final states, this should be multiplied by the QCD correction factor, $1+\alpha_s/\pi$.

\subsection{Results and constraints}\label{ssec:res}

\begin{figure}
  \centering
  \includegraphics[clip,width=0.49\textwidth]{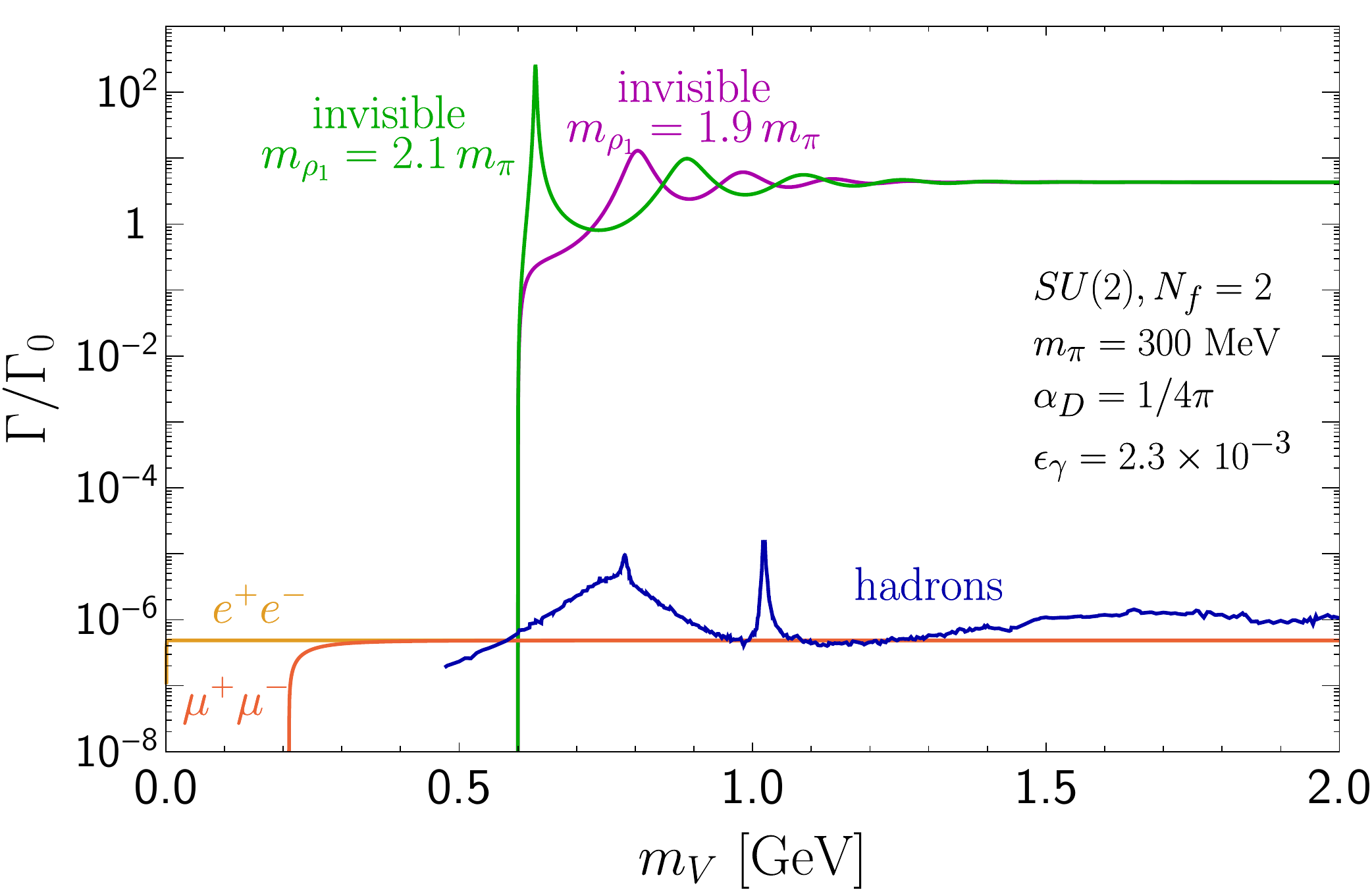}\hfill
  \includegraphics[clip,width=0.483\textwidth]{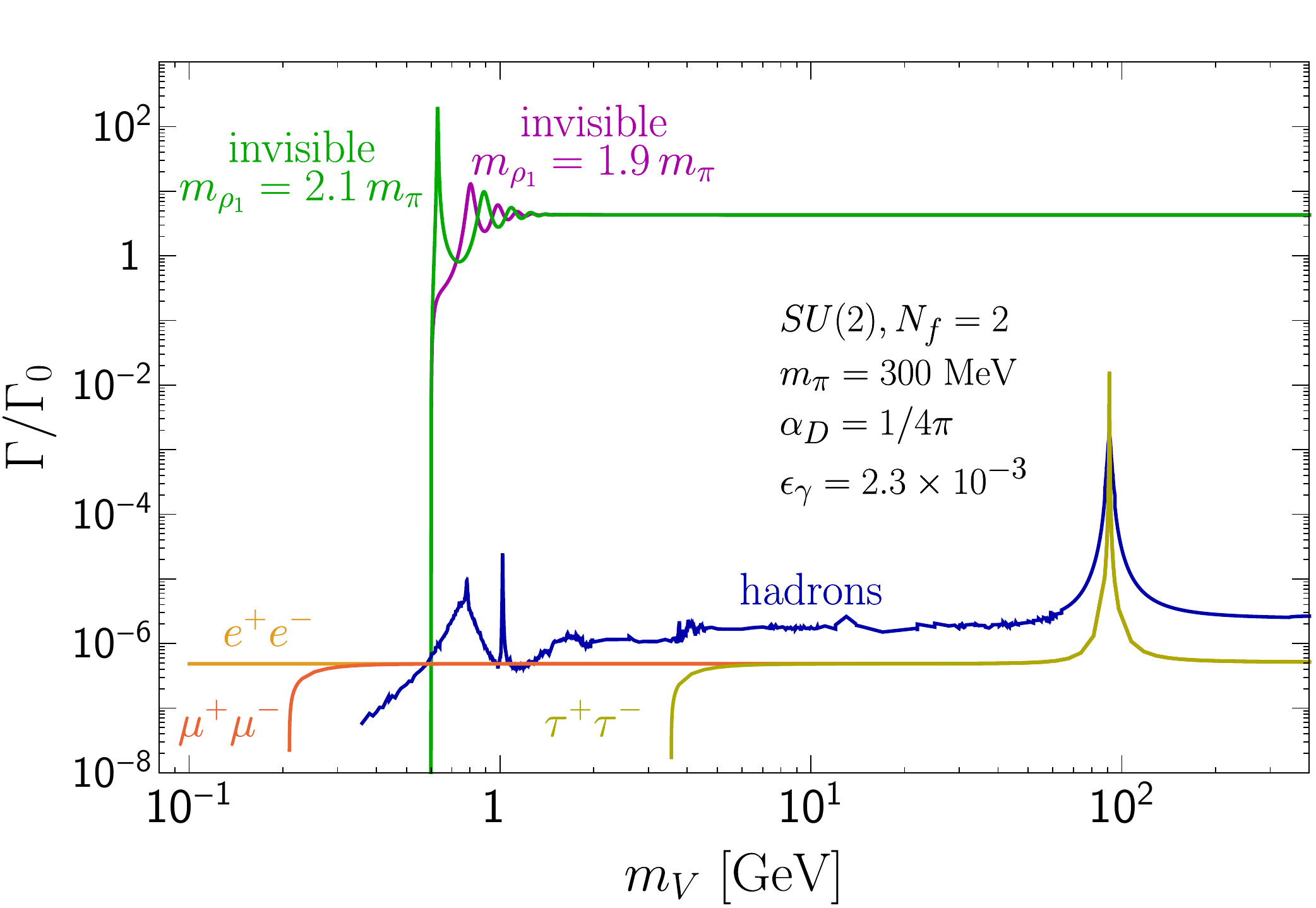}
  \caption{Partial widths of the dark photon into various final
    states, as a function of dark photon mass: invisible (purple
    and green), hadrons (blue), electrons (orange),
    muons (red) and taus (dark yellow, right panel only).
    We use the SIMPlest model of an $SU(2)$ gauge group with $N_f=2$
    (four doublets), with $m_\pi=300$~MeV, $\alpha_D = 1/(4\pi)$ and
    $\epsilon_\gamma = 2.3 \times 10^{-3}$.  The the partial width on
    the vertical axis is normalized to $\Gamma_0 = \alpha_D m_V/3$.  For
    the dark sector, we present two benchmark cases for $\rho$ masses:
    $m_{\rho_n}=1.9 m_\pi \sqrt{n}$~MeV (solid purple) and
    $2.1m_\pi \sqrt{n}$ (solid green), to illustrate the behavior
    of different possible dynamics. In the former case, the first
    state $\rho_1$ does not show up in the invisible width
    because it is lighter than $2m_\pi$; this is not the situation
    for the latter case. The right panel simply extends the mass range
    of the hidden photon to heavier masses than shown in the left
    panel.}
  \label{fig:R}
\end{figure}

We are now ready to compute the partial widths of the dark photon. If the dark photon is heavier than multi-GeV, it is in the perturbative regime for both our QCD and the QCD-like dark-sector.  For a light(er) dark photon, non-perturbative effects are important and the partial widths are highly non-trivial.

The last ingredient needed is the mass scale of the $\rho$ resonances
relative to $m_\pi$; this is highly model-dependent.  In the SIMPlest model of an $Sp(2)\simeq SU(2)$ gauge group with $N_f=2$, perturbativity considerations along with self-scattering constraints point to the rough scales of $m_\pi/f_\pi \approx 6$ and $m_\pi \approx 300$~MeV~\cite{Hochberg:2014kqa}. Since the theory is in the very strongly interacting regime, this is understood as a proxy to the scales involved; ${\cal O}(1)$ corrections are likely to apply. In what follows, we use these estimates for the pion scales involved for illustration purposes. We do not know what the corresponding $\rho$-meson mass is, however we do expect the first of the vector-meson states $\rho_n$ to at least be heavier than $\pi$.  This is because, in the non-relativistic quark model, the mass splitting between the mesons is due to the hyperfine interaction of two quarks inside the meson, which makes the $F=1$ state higher than the $F=0$ state.
The mass splitting narrows with increased quark mass. This is indeed the case in  QCD when going from $\rho$ and $\pi$, $\phi$ and $\eta$ to $J/\psi$ and $\eta_c$.  The vector meson is always heavier than the pseudoscalar. As a result, we will always take the $\rho$'s to be heavier than the pions.
Moreover, suppressing semi-annihilation processes of $\pi\pi\to \pi \rho$ (and in the SU($N_c$) case, also $\pi\pi\to \pi \omega$) compared to the $3\to2$ process requires $m_\rho \gsim 2m_\pi$.

\begin{figure}
  \centering
  \includegraphics[clip,width=0.8\textwidth]{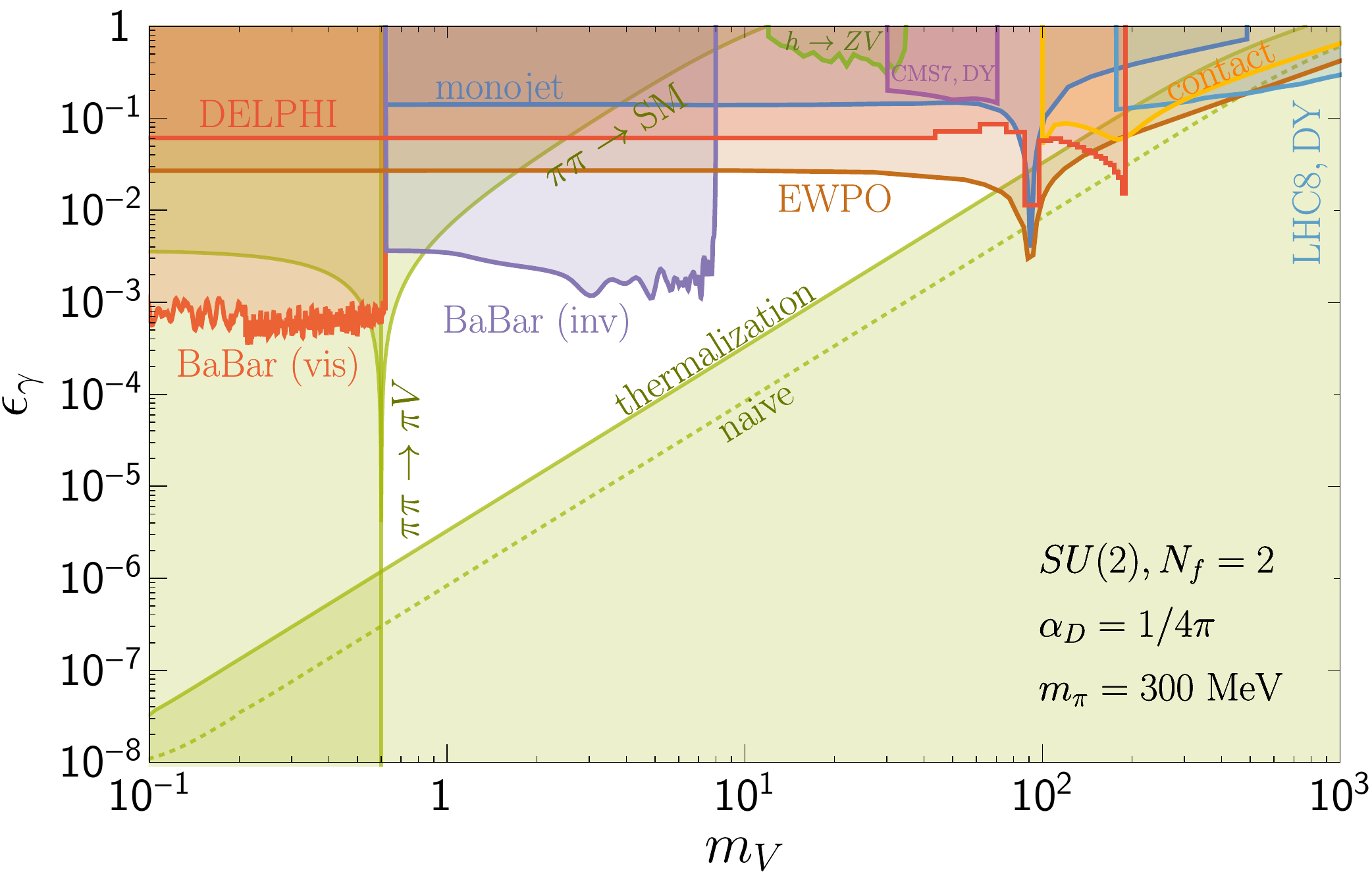}
  \caption{
  Combined requirements and constraints for SIMP dark matter with a dark photon mediator. The results are plotted for the SIMPlest model of an $SU(2)$ gauge group with $N_f=2$, $m_\pi=300$~MeV and $\alpha_D=1/(4\pi)$. The shaded green regions define the range of validity of the SIMP mechanism with a dark photon mediator to the visible sector, as in Fig.~\ref{fig:Thermalization}. The other colored shaded regions are the experimental constraints on $\epsilon_\gamma$, coming from visible~\cite{Lees:2014xha} and invisible~\cite{Aubert:2008as} decays at BaBar, mono-photon searches at DELPHI~\cite{Abdallah:2008aa}, electroweak precision observables~\cite{Essig:2013vha}, Drell-Yan production measured at ATLAS~\cite{Cline:2014dwa} and CMS~\cite{Chatrchyan:2013tia}, four-fermion contact operators at LEP II~\cite{Schael:2013ita}, $h\to ZV\to 4\ell$ from CMS data~\cite{Curtin:2014cca}, and monojet searches at CDF and ATLAS~\cite{Shoemaker:2011vi}.
  }
  \label{fig:SIMPlest}
\end{figure}

In Fig.~\ref{fig:R} we plot the various branching fractions of the dark photon as a function of its mass. The left and right panels of Fig.~\ref{fig:R} correspond to the low-mass and high-mass ranges of the dark photon, respectively. Here we take $m_\pi=300$~MeV, $\alpha_D = 1/(4\pi)$ and $\epsilon_\gamma = 2.3 \times 10^{-3}$, corresponding to the largest allowed value of $\epsilon_\gamma$ for the low mass range, as will be seen below in Fig.~\ref{fig:SIMPlest}. We choose two representative benchmark cases for lightest $\rho$-meson mass: $m_{\rho_1} = 1.9 m_\pi$ with the corresponding tower, and $m_{\rho_1}=2.1 m_\pi$ with the corresponding tower. Given that the $m_\rho/m_\pi$ mass ratio cannot be computed reliably, these two benchmark cases are sufficient as examples of different potential features. As mentioned above, we do not consider $\rho$ masses much below $2m_\pi$, such that semi-annihilations into $\rho$'s do not dominate the relic abundance, and are at most comparable to the $3\to2$ rate. If the mass of some low-lying vector state does not exceed $2m_\pi$, this resonance cannot decay invisibly and instead decays back to SM particles. This is exemplified by the $m_{\rho_1}=1.9m_\pi$ case, plotted in purple, where the first low-lying state decays into SM modes. Alternatively, the lightest low-lying vector state can be heavier than two pions, in which case all invisible modes are available. Such behavior is represented by the $m_{\rho_1}=2.1m_\pi$ case, which is plotted in green.

Having computed the partial widths, the dark photon constraints of Fig.~\ref{fig:DPlim} can be appropriately weighted and combined with the SIMP constraints of Fig.~\ref{fig:Thermalization} to obtain the viable parameter space. The results for the SIMPlest model of an $Sp(2)=SU(2)$ gauge group with $N_f=2$ are shown in Fig.~\ref{fig:SIMPlest}. The unshaded region in the center is phenomenologically viable. For instance, a dark photon with mass of a few GeV can have $\epsilon_\gamma$ as large as $\sim 10^{-3}$ but not smaller than $\sim 10^{-5}$; the $\epsilon_\gamma$ range mostly broadens (shrinks) for lighter (heavier) masses. Concerning the BaBar limits, our treatment is as follows. Given the smallness of the decay widths into SM particles compared to the decay width into dark resonances (see Fig.~\ref{fig:R}), we take 100\% branching fraction into SM particles when $m_V<2m_\pi$, and once decays into the hidden sector are kinematically allowed, we use 100\% branching fraction into the hidden sector. The depicted results use $m_\pi=300$~MeV and $\alpha_D=1/(4\pi)$, and the effect of altering the dark matter mass can be readily understood. As $m_\pi$ varies, the constraint from suppressing semi-annihilation $m_V\gsim 2m_\pi$ shifts accordingly, as does the resonant annihilation dip near $m_V=2m_\pi$. Likewise, the transition point between the BaBar SM final state searches and invisible modes will shift accordingly. Qualitatively, the viable parameter space of SIMPs shown in Fig.~\ref{fig:SIMPlest} well-captures the viable parameter space of other gauge groups with different number of colors and flavors as well.

\section{SIMP spectroscopy}\label{sec:spec}

If the dark-photon is produced off-shell, then its production probes the spectrum of resonances of the SIMP dynamics. The idea presented here applies more broadly to any strongly coupled sector that interacts in a similar way with the SM.

There are two distinct regimes to the SIMP physics that depend on whether the singlet dark $\rho$-meson is kinematically allowed to decay into the dark pions, or decays via the hidden-photon into SM states. All but the singlet $\rho$-mesons transform nontrivially under the broken flavor symmetry~(see Table \ref{tab:rho}), and may be stable against all decays, in the $Sp(N_c)$ and $SO(N_c)$ theories.  For the $SU(N_c)$ case, since the  $\rho$-mesons and pions transform identically, the $\rho$'s can decay into $\pi+V^{(*)}\rightarrow \pi+\ell^+ \ell^-$ etc.

When the dark $\rho$-meson can decay into dark pions, then any production of dark quarks or dark hadrons will quickly cascade decay into the dark pions, resulting in the dynamics of the dark sector being primarily measured indirectly. Since the SIMP parameter space naturally pushes $m_q/f_\pi > 1$~\cite{Hochberg:2014kqa}, and therefore $m_\pi \to m_\rho$, it is quite possible that the dark $\rho$-mesons will be stable against strong decays inside the dark sector.

If the dark $\rho$-meson is kinematically forbidden to decay into the
dark pions, then dark $\rho$-mesons must decay via interactions with
the dark photon. The singlet $\rho$-mesons can decay via
$\rho \rightarrow V^{(*)} \rightarrow \text{ SM}$. This can lead to a
lepton-jet-like signal \cite{Strassler:2006im}.

There are additional low-lying hadronic states that may play a role in
the phenomenology. The most important one is the $a_1$
resonance,\footnote{{Note that the subscript here refers to $J=1$, not
  the radial excitations $(a_1)_n$.  The same comment applies to
  $b_1$.}}  which is the $L=1$, $S=1$ spin-1 resonance. The $a_1$
transforms under the flavor symmetries as the pions. When the $a_1$
state is kinematically allowed to decay into the strong sector, it may
decay via the dominant processes $a_1\rightarrow 3\pi, \pi\rho_1$. If
the lowest $\rho$-mesons strong decays are forbidden, then strong
$a_1$ decays will be forbidden as well. Instead the $a_1$ will be
forced to decay via $a_1\rightarrow \pi V^{(*)}$. This decay is always
allowed, and so the $a_1$ will not be
stable.
Similar statements can be made about the $\omega$ resonances in
$SU(N_c)$ theories.

Additional states like the $L=1$, $S=0$ states ($b_1$ meson) are stable to decays to dark pions in $Sp(N_c)$ or $SO(N_c)$ theories. They will be able to decay with a dark photon into dark $\rho$-mesons. The number of states that are stable to strong decays grows as dark quark masses grow and the theory transitions into a  quirk-like regime~\cite{Kang:2008ea}.

\subsection{Measuring invisible mass}\label{ssec:Minv}

In $e^+ e^- \to \gamma+\rm inv$ annihilations, the energy of the recoil photon is in one-to-one correspondence with the invariant mass $M_{\rm inv}$ of the invisible system,
\begin{equation}
  E_\gamma = \frac{\sqrt{s}}{2} \left(1 - \frac{M_{\rm inv}^2}{s} \right)\,.
\end{equation}
The production cross section is given by
\begin{eqnarray}
  \sigma (e^+e^-\to \gamma+{\rm inv}) &=& \frac{3\alpha}{s} \int d \cos\theta \int dM_{\rm inv}^2
             \frac{M_{\rm inv}^2}{(M_{\rm inv}^2-m_V^2)^2+m_V^2\Gamma^2} \times \nonumber \\
         & &             \frac{\Gamma(V\rightarrow\mbox{inv})}{M_{\rm inv}}
             \frac{\Gamma(V\rightarrow e^+e^-)}{M_{\rm inv}}
             \frac{8-8\beta+3\beta^2+\beta^2\cos
             2\theta}{\beta\sin^2\theta}\,,
\label{eq:production}
\end{eqnarray}
with $\beta = 1-M_{\rm inv}^2/s$, which agrees with the result of Ref.~\cite{Berends:1987zz}.  Here, the widths are to be computed as if $m_V = M_{\rm inv}$, reflecting the off-shell nature of the dark photon in the process, using the $m_V$ dependence of widths given in Fig.~\ref{fig:R}.  We take the acceptance of photons to be  $|\!\cos\theta| < \cos 12^\circ$. This is only slightly optimistic compared to the actual geometric coverage of 12 to 157~degrees at Belle-II. Belle-II is working on implementing a single photon trigger~\cite{Higuchi}, {a significant effort considering the high rate of bremsstrahlung background}.

Using the partial widths modeled in Section~\ref{sec:decays}, we show the invariant-mass spectrum in $e^+ e^-$ collisions at $\sqrt{s}=10$~GeV. The entire mass range is shown in Fig.~\ref{fig:Minv}, while Fig.~\ref{fig:Minv2} focuses on the low mass range. The dark photon peak in the invariant mass distribution is clearly visible (here at 2~GeV). The irreducible background of $e^+ e^-  \to \gamma \nu \bar{\nu}$ is shown in the dashed gray curve in Fig.~\ref{fig:Minv}. For the low masses in Fig.~\ref{fig:Minv2}, we present two possible dynamics, using the benchmarks of $m_{\rho_1} = 1.9 m_\pi$ and $2.1 m_\pi$ as representative cases. In the latter case, when all $\rho$ resonances are above the two-pion threshold, their decays are entirely into the dark sector (green curve). In the former case, only the second and higher vector resonances decay invisibly (purple curve), while the first vector resonance cannot decay into dark pions, and so decays into charged leptons and QCD pions. We describe these visible decays next.

If the dark photon mass is well above the confinement scale, it
  decays into dark quarks which fragment dominantly into dark mesons.
  As discussed above, many of the meson states either decay into dark
  pions, are stable against the strong decays, or are completely stable.
  However, a fraction of the produced mesons can decay back into the Standard Model and
  create additional lepton pairs or QCD-hadrons.  The study of the
  invisible mass spectrum therefore needs to be done paying careful
  attention to this possibility that a fraction of the `invisible
  mass' is actually visible.

Finally, various dark sector masses can be measured without full knowledge of the strongly coupled sector. The invariant mass of lepton pairs would measure the mass splittings among the SIMP resonances, as well as the lowest $\rho$ states. The cutoff at small invariant mass in the $M_{\rm inv}$-distribution corresponds to $2m_\pi$, measuring the mass of the dark matter.  A combination of these techniques can in principle reveal the spectrum of the SIMP sector resonances.

{\begin{figure}[t!]
  \centering
  \includegraphics[clip,width=0.77\textwidth]{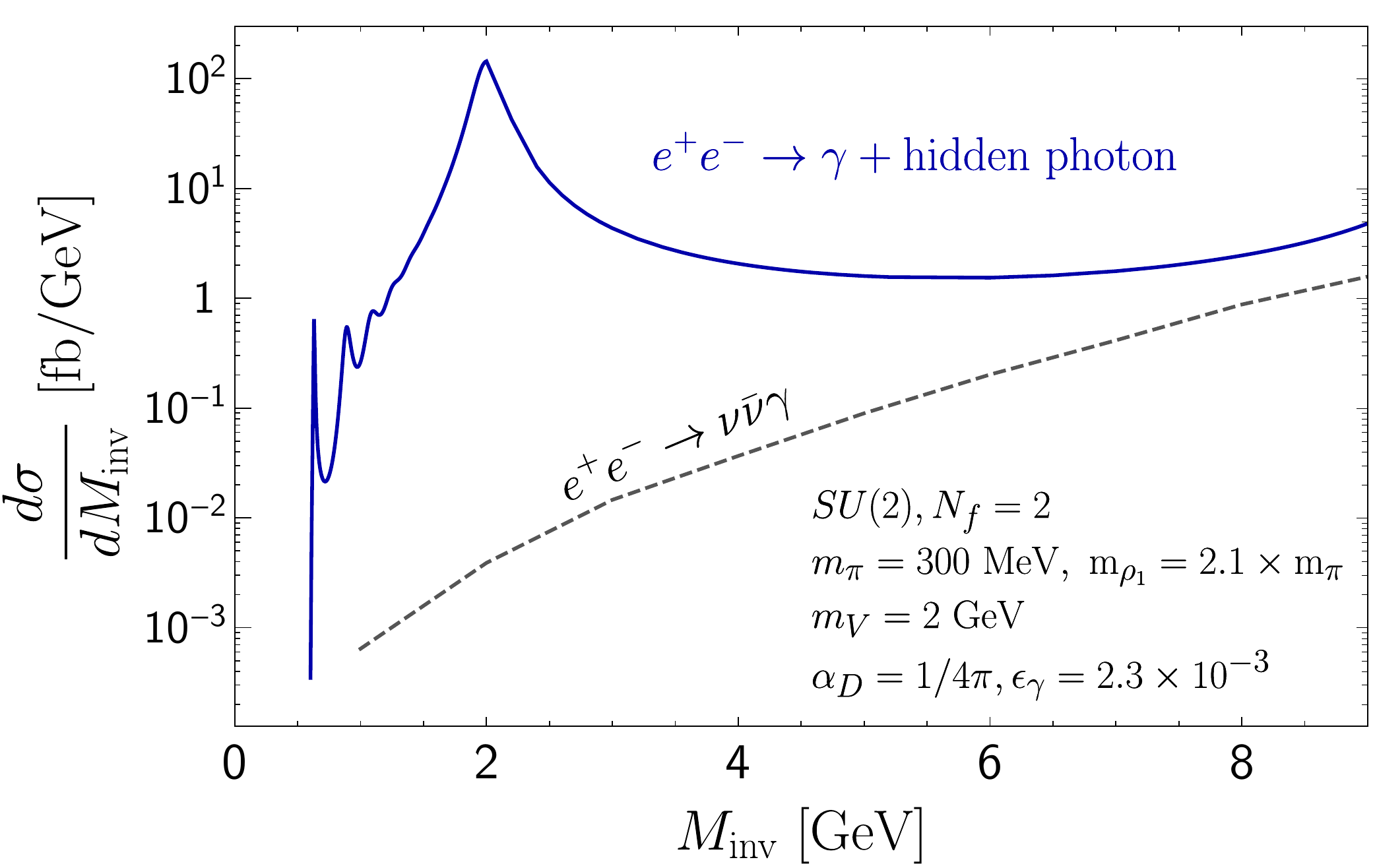}
  \caption{The cross section for
    $e^+ e^- \rightarrow \gamma+\mbox{dark photon}$ at
    $\sqrt{s} = 10$GeV, where the dark photon decays invisibly (solid blue curve). We plot this
    cross section in fb/GeV as a function of the invariant mass of the invisible system.
    The dashed gray line is the irreducible background coming from
    $e^+ e^- \rightarrow \gamma+Z^*$ with $Z^* \rightarrow \nu \bar{\nu}$.
    Here we take the SIMPlest model of an $SU(2)$ gauge group and $N_f=2$
    (four doublets), with $m_\pi=300$~MeV, $m_{\rho_n}=2.1m_\pi \sqrt{n}$~MeV, $\alpha_D = 1/(4\pi)$ and
    $\epsilon_\gamma = 2.3 \times 10^{-3}$.  We use a dark photon mass of 2~GeV, and its peak is clearly visible in the $M_{\rm inv}$ spectrum.   \label{fig:Minv}}

\vspace{.8cm}
  \centering
  \includegraphics[clip,width=0.77\textwidth]{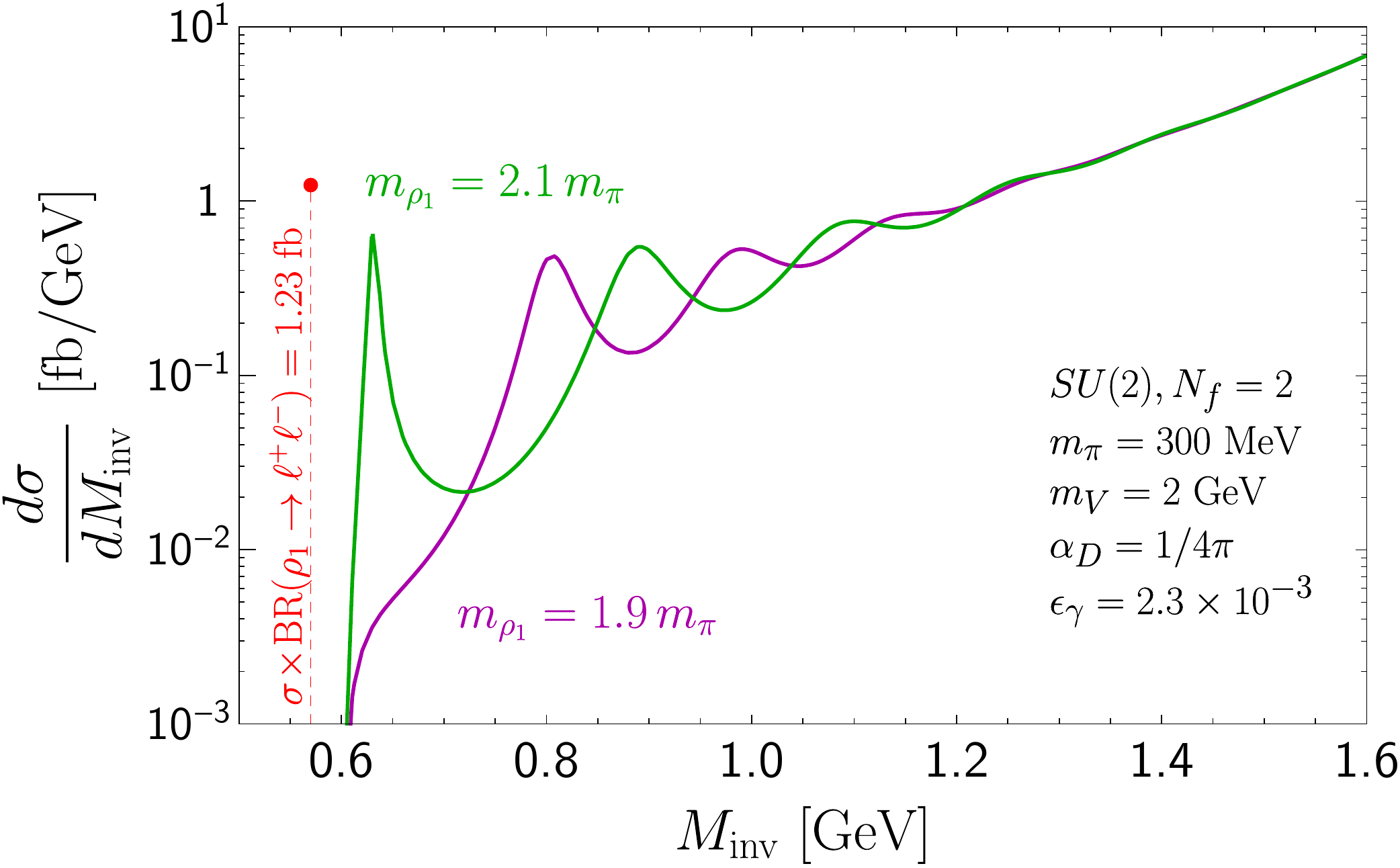}
  \caption{The same as in Fig.~\ref{fig:Minv}, but focusing on the
    low invariant mass region. Two benchmarks are considered. When $m_{\rho_1} = 2.1 m_\pi$ (solid green), all
    $\rho$ states decay into $\pi\pi$.  When $m_{\rho_1} = 1.9 m_\pi$ (solid purple),
    the first $\rho$ state cannot decay invisibly and decays instead into charged leptons. These decays are described in the dashed red vertical line, and the corresponding $\sigma\times{\rm BR}$ is marked as well.  \label{fig:Minv2} }

\end{figure}}

\subsection{Visible widths of SIMP resonances}\label{ssec:visSIMP}

If a resonance in the dark sector cannot decay `strongly', say when $m_\rho < 2 m_\pi$, it will decay into the visible sector.  This is akin to the fact that $J/\psi$ cannot decay into $D^+ D^-$, and decays instead into $\mu^+ \mu^-$ with a large branching fraction ($\sim6\%$).  In the case of $SU(N_c)$ gauge theories, $\omega$ cannot decay into two pions but can into three pions.  Therefore the $\omega$ states are more likely to decay into the Standard Model leptons.

The leptonic branching fraction for a dark resonance can be computed in the same way as for the vector mesons of ordinary QCD. For the ordinary $\rho$-meson of the SM,
\begin{eqnarray}
  \langle 0 | J_{EM}^\mu | \rho_{\rm SM}\rangle
  &=& \langle 0 | e \left(\frac{2}{3} \bar{u}\gamma^\mu u
    - \frac{1}{3} \bar{d} \gamma^\mu d \right) | \rho_{\rm SM} \rangle
= e \langle 0 | \frac{1}{2} (\bar{u} \gamma^\mu u - \bar{d} \gamma^\mu d
  )  | \rho_{\rm SM} \rangle \nonumber \\
  &=& e F_{\rho_{\rm SM}} \epsilon^\mu\,.
\end{eqnarray}
As a result, for the SM $\rho$-meson
\begin{equation}
    \Gamma(\rho_{\rm SM} \rightarrow e^+ e^-) = \frac{4\pi}{3}
  \frac{\alpha^2F_{\rho_{\rm SM}}^2}{m_{\rho_{\rm SM}}^3}\,.
\end{equation}

Following the SM example, but for the case of $Sp(N_c)$ theories and
the singlet $\rho$-mesons (the vector corresponding to the gauge
$U(1)_D$-symmetry, up to a normalization),
\begin{equation}
  \langle 0 | J_D^\mu | \rho_n\rangle
  = \langle 0 | e_D \sum_{i=1}^{N_f} \bar{q}_i \gamma^\mu q_i  | \rho_n \rangle
  = e_D \sqrt{2N_f} F_{\rho_n} \epsilon^\mu\,.
\end{equation}
Here, $q_i$ are regarded as Dirac fermions.
We take the decay constants from the soft wall model~\cite{Karch:2006pv},
\begin{equation}
  F_{\rho_n}^2 = \frac{m^2_{\rho_n} m^2_{\rho_1} N_c}{24 \pi^2}\,,
\end{equation}
see Appendix~\ref{sssec:holo} for more details. Taking into account the mixing between the dark photon and our
photon, and considering the $SU(N_c)$ and $SO(N_c)$ gauge theories as well, the widths into dilepton final states are\footnote{We thank Asher Berlin and Nikita Blinov for pointing out a mistake in the width in a previous version.}
\begin{equation}
        \Gamma(\rho_n \rightarrow \ell^+ \ell^-) =
  \frac{\alpha_D \epsilon_\gamma^2 \alpha}{18\pi}\frac{m_{\rho_n}^4}{m_V^4}
  \frac{m_{\rho_1}^2}{m_{\rho_n}}
  \frac{ \beta(3-\beta^2)}{2} N_c N_f^{\it eff},
      \label{eq:Gammaee}
\end{equation}
where
\begin{equation}
  N_f^{\it eff} =  \left\{\begin{array}{lcc}
2N_f & & Sp(N_c)\\
N_f & & SO(N_c)\\
{\frac{8N_1N_2}{N_1+N_2}} & & SU(N_c)
\end{array}\right.\, .
\end{equation}
 In the case of $SU(N_f)$ theories, there is also the $\omega_n$, which is
degenerate with the $\rho_n$ in holographic QCD, with width
\beq
\Gamma(\omega_n \rightarrow e^+ e^-) &=&
  \frac{\alpha_D \epsilon_\gamma^2 \alpha}{18 \pi }\frac{m_{\rho_n}^4}{m_V^4}
  \frac{m_{\rho_1}^2}{m_{\rho_n}}
  \frac{ \beta(3-\beta^2)}{2} {N_c \frac{2 (N_1-N_2)^2}{N_f}}\,.
\eeq
Finally, there are widths into (our) hadrons, given by the
multiplication of the above expressions by $R(s)$ of Eq.~\eqref{eq:Rs} with $\beta=1$.

We can use the production cross section in Eq.~(\ref{eq:production}) by identifying the vector $V$ not as the dark photon but rather as SIMP resonances, in the narrow width approximation:
\begin{equation}
  \int dM^2 \frac{1}{(M^2-m^2)^2+m^2\Gamma^2}
  = \frac{\pi}{m \Gamma}\, .
\end{equation}
Then Eq.~(\ref{eq:production}) reduces to
\beq
\lefteqn{  \frac{d\sigma(e^+e^- \rightarrow\gamma\rho \rightarrow \gamma
    \ell^+ \ell^-)}{d\cos\theta} } \nonumber \\
  &=&
             \frac{\alpha^2 \epsilon_\gamma^2  \alpha_D }{4 s}  \frac{m_{\rho_1}^2}{m_{\rho_n}^2}
             \frac{8-8\bar{\beta}+3\bar{\beta}^2+\bar{\beta}^2\cos
             2\theta}{\bar{\beta}\sin^2\theta}
             {\rm BR}(\rho_n\rightarrow \ell^+ \ell^-) N_c  N_f^{\it eff},
\eeq
with $\bar{\beta} = 1-m_{\rho_n}^2/s$. The results are shown in
Fig.~\ref{fig:Minv2} in the red vertical line, for the
first vector resonance of the $m_{\rho_n}=1.9 m_\pi n^{1/2}$
tower for the $SU(2)$ gauge theory with $N_f=2$.

\section{Future prospects}\label{sec:future}

We now discuss future prospects for SIMP dark matter searches interacting with the SM via a kinetically mixed hidden photon. These include direct detection, spectroscopy at $e^+e^-$ colliders, hadron collider signatures, ILC prospects and beam dump experiments. To illustrate the many future probes of the setup, in Fig.~\ref{fig:SIMPfuture} we show projections of various future measurements into the viable parameter space of Fig.~\ref{fig:SIMPlest}. The shaded green and gray regions correspond to the excluded regions of Figs.~\ref{fig:Thermalization} and~\ref{fig:SIMPlest}, accordingly, while the colored curves show future reach, as detailed below.

\subsection{Dark matter direct detection}\label{ssec:DD}

Sub-GeV dark matter is challenging for direct detection
experiments relying on conventional nuclear recoil signals (though see Refs.~\cite{Agnese:2013jaa,Agnese:2014aze,Agnese:2015nto}).
For  light dark matter of order hundreds of MeV, scattering off an electron will typically deposit $m_e v_{\rm halo}^2 \simeq 100\; \mbox{eV}$ energy, and can excite
electrons in a semiconductor above the band gap, leading to a
detectable signal~\cite{Essig:2011nj,Graham:2012su}. (In this context, see Ref.~\cite{Hochberg:2015pha,Hochberg:2015fth} for use of superconducting targets as well.)

The non-relativistic scattering cross section of SIMP pions on electrons is given by
\begin{equation}
  \sigma =  \frac{1}{8\pi}
  \left( \frac{2 \epsilon_\gamma e e_D}{m_V^2} \right)^2 s \bar{\beta}^2\,, ~~~~~   \bar{\beta} = 1 - \frac{m_\pi^2}{s}\,.
\end{equation}
Here, $s=(m_e+m_\pi)^2 + m_e m_\pi v_\pi^2$.

\begin{figure}
  \centering
   \includegraphics[clip,width=0.8\textwidth]{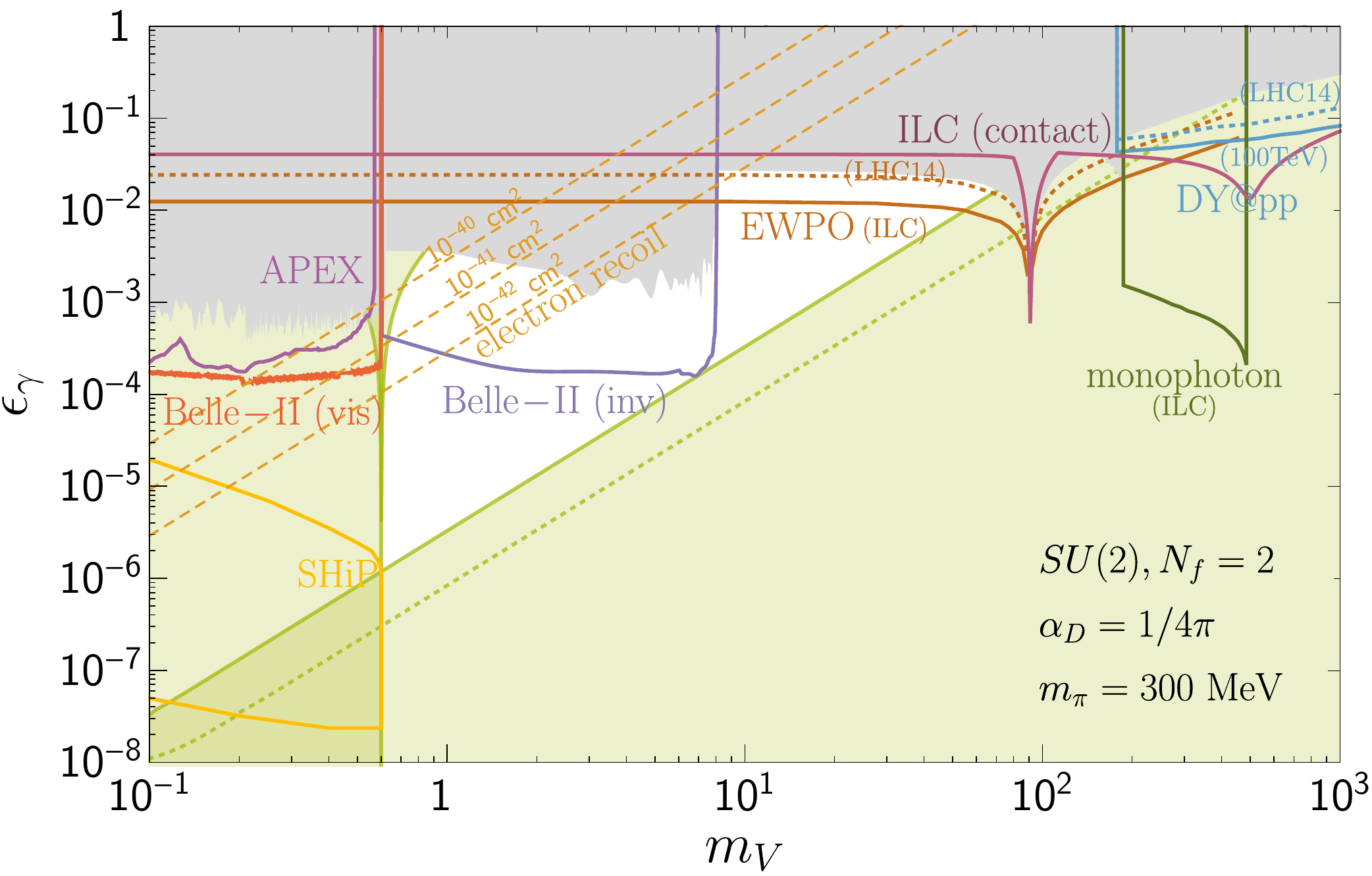}
  \caption{Anticipated future reach of various experiments into the viable parameter space of the SIMP/dark photon setup. These are:
    Belle-II visible~\cite{Soffer:2014ona} (solid red) and invisible~\cite{Essig:2013vha} (solid purple); SHiP~\cite{Alekhin:2015byh} (solid orange); Drell-Yan at LHC14~\cite{Curtin:2014cca} (dashed blue) and 100~TeV $pp$ collider~\cite{Curtin:2014cca} (solid blue); ILC monophoton search~\cite{Titov} (solid green), reinterpretation of contact operators at the ILC~\cite{Djouadi:2007ik,Riemann} (solid violet) and EWPO at the LHC (dashed brown) and the ILC~\cite{Essig:2013vha,Curtin:2014cca} (solid brown); direct detection via electron recoil signals (dashed orange); and APEX~\cite{Essig:2010xa} (solid dark purple).
    }
      \label{fig:SIMPfuture}
\end{figure}

For dark matter mass of order a few hundred MeV, the current direct detection constraint is $\sigma_{\rm DD} \lesssim
10^{-35}~\mbox{cm}^2$~\cite{Essig:2012yx} using data from the Xenon10 experiment~\cite{Angle:2011th}.  The analysis of Ref.~\cite{Essig:2015cda} suggests that an
improved reach down to cross sections of ${\cal O}(10^{-41}~\mbox{cm}^2)$ may be achievable in a silicon or germanium semiconductor detector with a
kg$\cdot$year of exposure.  Contours of the SIMP-electron elastic scattering cross section are overlaid in
Fig.~\ref{fig:SIMPfuture} in dashed orange curves.

\subsection{Spectroscopy at low-energy lepton colliders}

At SuperKEKB, an integrated luminosity of 50~ab$^{-1}$ is
anticipated.
In the Belle II experiment, the implementation of a single photon trigger has started to be studied in order to make the beam-induced and QED-originated backgrounds under control.
The expected energy resolution of the electromagnetic calorimeter is 1.5\% at
$E_\gamma=5$~GeV, 2\% at $E_\gamma=1$~GeV and 4\% for
$E_\gamma = 0.1$~GeV~\cite{Higuchi} (see also Ref.~\cite{Bevan:2014iga}).  There is a large background at
$M_{\rm inv}=0$ when one of the photons in the $e^+ e^- \rightarrow \gamma\gamma$ process is lost.
This should be estimated by a detector Monte Carlo, and one can also veto events where a single photon is detected whose recoil photon falls into the insensitive region of the detector with the $e^+ e^- \rightarrow \gamma\gamma$ hypothesis~\cite{Higuchi}.  With the expected energy resolution, Belle-II should, in principle, be able to separate individual
peaks of a resonance structure from each other as well as from the
background.  They could also find the dark photon resonance itself and
possibly multiple peaks in lepton pairs, leading to a striking signal, as shown in Figs.~\ref{fig:Minv} and~\ref{fig:Minv2}.

If a signal is seen, a dedicated low-energy high-intensity $e^+ e^-$ collider would enable more detailed spectroscopy to be performed.
From the lattice perspective, it may become possible in the future to compute the vector-vector
correlation function for time-like momenta by analytic continuation
from the precise lattice calculations~\cite{Fodor}, in which case one could
extract many characteristics of the strongly coupled hidden sector, such as the gauge group, number of colors, number of flavors and the quark masses, by fitting the spectrum.

\subsection{LHC prospects}

Most of the current constraints on the dark photon parameter space from the LHC are expected to improve in the near
future, via the search for the Drell--Yan process, search
for monojet events, and more detailed studies in $h \rightarrow 4\ell$
at the LHC Run-2 and beyond. Projections for the LHC and at a 100~TeV $pp$ collider are shown in Fig.~\ref{fig:SIMPfuture}. The 100 TeV Drell-Yan and  EWPO LHC14 projections are taken from~\cite{Curtin:2014cca}.

Other interesting LHC phenomenology is expected.  At the LHC, in the entire allowed parameter space, the dark photon can be produced on-shell and often with a significant amount of $p_T$. If the dark photon mass is significantly above the confinement scale, then dark photon decays produce dark quarks. The dark quarks will shower and fragment dominantly into dark mesons.
If $m_\rho/m_\pi \sim 1$, there is no kinematical suppression for fragmenting into the dark $\rho$ mesons, and dark $\rho$ mesons may become the dominant final states {due their three spin states}. If  $m_\rho < 2m_\pi$, then for the $SO(N_c)$ and $Sp(N_c)$ cases, the singlet dark $\rho$-meson can decay into lepton pairs, while the adjoint ones into $\pi+V^{(*)} \rightarrow \pi+\ell^+ \ell^-$.  The remaining ($\Ysymm$ and $\Yasymm$) dark $\rho$-mesons are stable. Thus, the signals in this case will primarily be a mix of missing energy and $\rho$-decays via $V^{(*)}$ into narrowly collimated
small invariant mass lepton pairs, known as `lepton jets'~\cite{ArkaniHamed:2008qp,Baumgart:2009tn,Bai:2009it,Cheung:2009su}. The invariant mass of the lepton pairs could be used to reconstruct the SIMP resonances. The primary challenge in detecting events where $pp \rightarrow V + X$ is that the lepton jets will be relatively soft because the number of dark hadrons that are typically produced in the fragmentation is large. Depending on the size of the kinetic mixing, $V$ may decay promptly,
or  displaced from the collision point. This is an explicit realization of `hidden
valley'~\cite{Strassler:2006im} phenomenology. If the hidden QCD scale (and therefore the dark matter mass) is heavier, emerging jets~\cite{Schwaller:2015gea} are possible as well.

Additional states like the $L=1$, $S=1$ dark-$a_1$ state, or even
possibly the $L=1$, $S=0$ states, can be produced in the
fragmentation of the dark quarks.  These decays will produce
$V^{(*)}$ signals. We cannot predict from first
principles the rate with which that these high resonances appear in the dark
quark jets, though they are likely to be produced at lower rates since they will be
heavier than the $\pi$ and $\rho$ states. However, if they are produced at
significant numbers, then the production of rich lepton-jet events is
possible.

\subsection{ILC prospects}

We study the ILC sensitivity using two processes. The first is the search
for monophoton events, similar to the search discussed earlier at LEP II with
DELPHI data.  The second is precision measurements of $e^+ e^-
\rightarrow f \bar{f}$ processes, which are usually cast in the language of
contact (effective four-fermion) operators.

The monophoton search is limited by the background.  We rely on the
simulated background levels in Ref.~\cite{Titov}, assuming
$\sqrt{s} = 500$~GeV, with integrated luminosity of 500~fb$^{-1}$
and beam polarizations $P_{e^-}=0.8$, $P_{e^+}=-0.6$.  (This polarization suppresses the background as the electron is nearly right-handed and does not couple to the $W$ boson.) We use the
narrow width approximation, so that the dark photon production appears
in one bin of the recoil photon energy, and allow for $2\sigma$
fluctuations relative to the simulated background.  The sensitivity
goes far beyond the current limits, as is shown in the green curve in Fig.~\ref{fig:SIMPfuture}.  Unfortunately the simulation in Ref.~\cite{Titov} does not extend to photon energies above $E_\gamma > 220$~GeV, so
the sensitivity for $m_V < 173$~GeV cannot be evaluated.  We hope that studies will be
conducted for smaller dark photon masses as well.

Contact operator analyses are the standard analyses in
$e^+ e^- \rightarrow f \bar{f}$ processes via the interference with
the Standard Model amplitudes, and are expressed by the energy scale
of the dimension-six four-fermion operators.  Compared to the LEP II
limits of approximately 20~TeV \cite{Schael:2013ita}, the ILC is expected to probe the energy scale up to
nearly 200~TeV~\cite{Djouadi:2007ik}, based on Ref.~\cite{Riemann}.  Since
the center-of-mass energy is fixed at the ILC, unlike the LHC case, it is straightforward
to re-interpret the ILC sensitivity to the case where the dark photon mass is
lower than the center-of-momentum energy. The resulting projected reach is shown in the violet curve in Fig.~\ref{fig:SIMPfuture}.

In addition, the ILC will tighten EWPO constraints with its GigaZ
option~\cite{Essig:2013vha,Curtin:2014cca}, as shown in the solid brown curve in Fig.~\ref{fig:SIMPfuture}.

\subsection{Beam dump and fixed target experiments}

If the dark photon decays visibly, it can be searched for in beam
dump experiments.  When a proton beam hits a thick target, many hadrons are produced, some of which may decay into the dark photon.
The dark photon may additionally be produced via bremsstrahlung off a proton. After all the hadrons and muons are stopped and/or swept
away, only very weakly interacting particles will propagate into the
detector.  If the dark photon decays into $\ell^+ \ell^-$ or hadrons,
it can be detected in this fashion.

Since the dark photon branching fraction into the visible particles is
extremely small if it can decay to the strong sector, the sensitivity of beam dump experiments to the dark photon is
limited to $m_V < 2m_\pi$. However, such a search would be very sensitive to
small values of $\epsilon_\gamma$. In the orange solid curve of Fig.~\ref{fig:SIMPfuture}, we show the projected reach of the SHiP
(Search for Hidden Particles) facility, proposed for the CERN SPS~\cite{Alekhin:2015byh}, into the dark photon parameter space.

Finally, the projected reach for the proposed electron fixed-target experiment APEX~\cite{Essig:2010xa} is shown in the solid dark purple curve of Fig.~\ref{fig:SIMPfuture}.

\section{Discussion}\label{sec:conc}

In this paper, we discussed the phenomenology of strongly interacting
massive particle (SIMP) dark matter.  We use a dark photon that is kinetically mixed
with $U(1)_Y$ to maintain kinetic equilibrium with the Standard Model
sector during freeze-out.  We studied a host of experimental constraints on the dark photon and its coupling
to the dark matter.  A  wide parameter
range is allowed by all constraints, summarized in Fig.~\ref{fig:SIMPlest}.
Stringent limits on dark matter annihilations at the time of recombination as well as
from indirect-detection are evaded due to the $p$-wave nature of the SIMPs
annihilation through the dark photon exchange.

Many current and forthcoming experiments will improve their
sensitivity to the remaining viable parameter space, as is summarized in Fig.~\ref{fig:SIMPfuture}. The novel idea proposed here is to
use the dark photon to study the spectroscopy of the vector meson resonances in the SIMP sector.  At an
$e^+ e^-$ collider, tagging the photon and measuring its energy uniquely
determines the recoiling system, even when it is invisible.  Alternatively, some resonances may decay back into Standard Model
leptons, if their mass is too light to decay into the SIMP sector.  The dark photon peak could be clearly seen in the invariant mass distribution of the mono-photon events, and the resonant structure of the strongly coupled theory could be detected, as is illustrated in Figs.~\ref{fig:Minv} and~\ref{fig:Minv2}. A single-photon trigger at Belle-II could prove crucial in exploring this dark spectroscopy. The
spectroscopy would allow us in principle to determine characteristics of the dark sector such---as the gauge group,
the particle content, and the quark masses in the SIMP sector---once a
better theoretical handle to compute dynamics of strongly coupled
gauge theories is at hand. More generally, such spectroscopy can be use to study any strongly coupled sector that couples to the Standard Model in a conceptually similar way, even if not related to dark matter. At the LHC, the (on-shell or off-shell) dark photon
can produce dark quarks which fragment into some of the low-lying
resonances which decay into Standard Model leptons or jets, leading to lepton-jet and/or emerging-jet signatures.  The direct detection of SIMP dark matter appears possible using semi-conductor detectors with electron recoil signals in parts of the viable parameter space.

\acknowledgments

We thank Zolt\'{a}n Fodor, Tongyan Lin, Takeo Higuchi, Kenkichi
Miyabayashi, Yu-Dai Tsai for useful discussions, and Tomer Volansky  and Jay Wacker for helpful conversations and comments on the manuscript. We thank the journal referee for an important comment on thermal tails. We are especially grateful to Jay Wacker for contributions to parts of this manuscript.  The work of YH is supported by the U.S. National Science Foundation under Grant
No. PHY-1002399. YH is an Awardee of the Weizmann Institute of Science
-- National Postdoctoral Award Program for Advancing Women in
Science. EK is supported by the NSF under Grant No. PHY-1316222 and  by a Hans Bethe Postdoctoral Fellowship at Cornell.  HM
was supported by the U.S. DOE under Contract DE-AC02-05CH11231, and by
the NSF under grants PHY-1002399 and PHY-1316783.  HM was also
supported by the JSPS Grant-in-Aid for Scientific Research (C)
(No.~26400241), Scientific Research on Innovative Areas
(No.~15H05887), and by WPI, MEXT, Japan.  This work
was supported in part by the NSF under Grant No. PHYS-1066293 and the
hospitality of the Aspen Center for Physics.

\appendix

\section{Hadronic widths from holographic QCD}\label{sssec:holo}

In Section~\ref{ssec:resonances}, the main results of this appendix were presented.

\begin{figure}[t!]
  \centering
  \includegraphics[clip,width=1\textwidth]{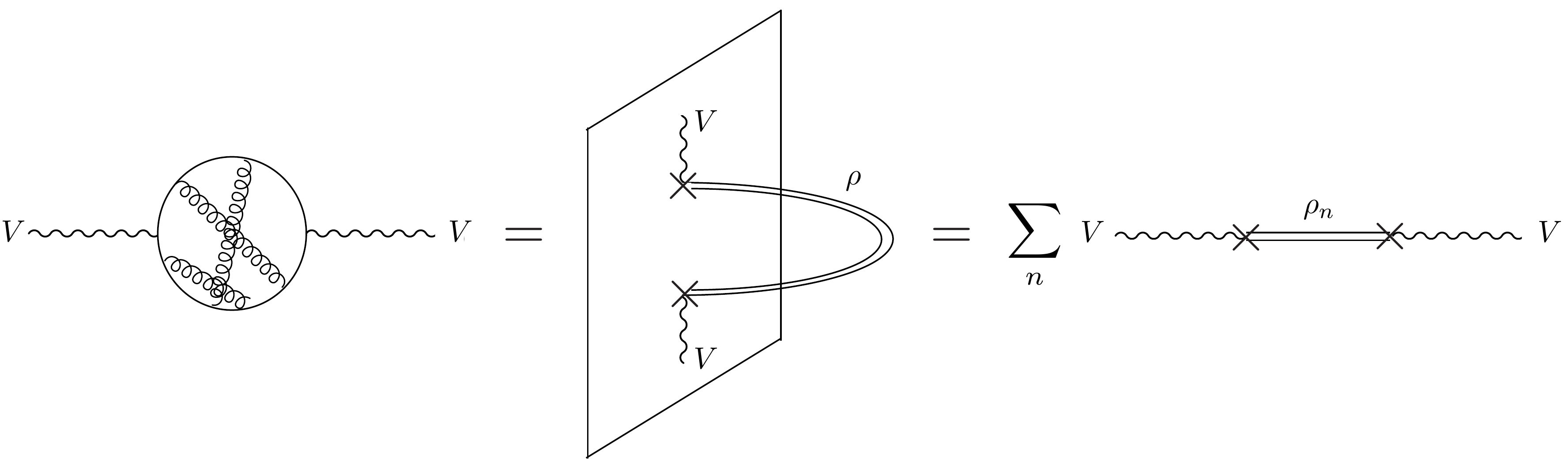}
  \caption{Schematic illustration of holographic QCD.  The left most diagram
    represents the vacuum polarization diagram in the strongly coupled
    QCD-like gauge theory, where the external line is the dark
    photon.  Using holography, it is computed in AdS$_5$ where the
    dark photon lives on the brane while the gauge boson of the flavor symmetry extends
    into the bulk.  Using the KK expansion of the bulk gauge boson, it
    is replaced by the sum over the KK tower in four dimensions.
    \label{ADSQCD}}
\end{figure}

We choose to model the resonance structure using the idea of
holographic QCD (see Ref.~\cite{Erlich:2014yha} for a review).  The
idea is as follows.  The flavor symmetry in the dark QCD-like sector
is promoted to a gauge symmetry in the AdS bulk.  A scalar field $X$
in the bulk, with the same quantum numbers as the quark bilinear
condensate, acquires an expectation value and spontaneously breaks the
flavor symmetry. The gauge fields have Kaluza--Klein (KK) towers that
correspond to the radial excitations of the spin-one bound states.
The broken gauge fields are the analog of the $a_1$ axial-vector
mesons of QCD, while the unbroken gauge fields, which also acquire
mass due to the boundary conditions~\cite{Csaki:2003dt}, are the
analog of the $\rho$ vector mesons of QCD.  The quantum numbers of the
$\rho$-mesons can be read off from gauging $H$, which matches with the
intuition from the quark model with $L=0$, $S=1$ bound states; see
Table~\ref{tab:rho}.  In the case of $SU(N_c)$ gauge theories, there
is an additional $U(1)_B$ gauge boson that is the analog of the
$\omega$ vector mesons of QCD. A schematic illustration of the setup
is given in Fig.~\ref{ADSQCD}. A reader interested directly in the
results of the invisible width modeling can skip to the summary given
in the final paragraph of this appendix.

Given that we normalize the generators of the flavor symmetry as
${\rm Tr} T^a T^b = 2 \delta^{ab}$~\cite{Hochberg:2014kqa}, as opposed
to the more conventional ${\rm Tr} T^a T^b = \frac{1}{2} \delta^{ab}$,
we define the covariant derivative for the bulk gauge field to be
  \begin{equation}
    D_\mu = \partial_\mu - i \frac{1}{2} g_5 T^a A_\mu
  \end{equation}
so that in what follows, the definition of the gauge coupling constant $g_5$ matches that in the literature. Correspondingly, the Noether currents are
\begin{equation}
  J_\mu^a = \frac{1}{2} q^\dagger \gamma_\mu T^a q\,.
\end{equation}

The spectrum of the unbroken gauge sector is relatively
model-independent, fixed only by the boundary conditions of AdS.  An
early attempt in the literature using the `hard wall' boundary
produced a spectrum $m_n^2 \sim n^2$~\cite{Erlich:2005qh}, which does
not match the observed Regge behavior.  Instead, we use here the
spectrum obtained using the `soft wall' model that reproduces the
Regge behavior, $m^2_n \propto (n+J)$~\cite{Karch:2006pv} (details
below). In contrast, the spectrum of the broken gauge sector, namely
that of $a_1$ and $\pi$ in the case of QCD, is highly model-dependent
as it depends on the potential for the field $X$ in the bulk and its
interactions.  In the models discussed in this work, the dark photon
does not mix into axial vectors, and these are omitted from further
discussion in this appendix.

The key point is that in holographic QCD, the sum over the KK towers
in the vector-vector two-point function reproduces the parton-level
vacuum polarization diagram in the large $N_c$ limit.  Here and below,
we only consider the vector currents of the unbroken symmetries.  The
definition of the vector-vector correlation
function~\cite{Erlich:2005qh} is
\begin{equation}
  \int d^4x e^{i q\cdot x} \langle J_\mu^a(x) J_\nu^b(0) \rangle
  = i \delta^{ab} (q_\mu q_\nu - g_{\mu\nu}) \Pi(q^2).
\end{equation}
Holography states that this is equivalent to the propagation of the
gauge boson in the bulk.  By writing the gauge boson in terms of its
KK states, with the `soft-wall' background in AdS~\cite{Karch:2006pv},
the KK spectrum of the unbroken gauge bosons is simply
\begin{equation}
  m_n^2 = 4 n\,, \qquad n = 1,~2,~\ldots
  \label{eq:mn}
\end{equation}
in units of the AdS curvature. The KK gauge field wave functions are
given in terms of the associated Laguerre polynomials,
\begin{equation}
  v_{n+1} (z) = z^2 \sqrt{\frac{2n!}{(n+1)!}} L_n^1 (z^2)\,.
\end{equation}
Then the vacuum polarization diagram can be obtained by the sum over
the KK tower of the gauge bosons,
\begin{equation}
  \Pi(q^2) = - \frac{1}{g_5^2} \sum_n
  \frac{(v^{\prime\prime}_n (0))^2}{(q^2-m_n^2+i\epsilon)m_n^2},
  = - \frac{1}{g_5^2} \sum_n \frac{2}{q^2-4n+i\epsilon}\ .
\end{equation}
The sum over the tower diverges logarithmically as expected, but is
renormalized,
\begin{equation}\label{eq:PiHarmonic}
  \Pi(q^2)-\Pi(0)
  = - \frac{2}{g_5^2}
  \sum_n \frac{q^2}{(q^2-4n+i\epsilon)4n} = - \frac{2}{g_5^2}
  \frac{1}{4}H_{-q^2/4}\, ,
\end{equation}
where $H_r$ is the harmonic number of order $r$.  Comparing it to
Eq.~(\ref{eq:Pi}), we can read off
\begin{equation}\label{eq:Fn}
  \frac{F_{\rho_n}^2}{m_{\rho_n}^2} = \frac{2}{g_5^2}\ .
\end{equation}

The overall normalization, determined by $g_5$, is still
undetermined. Using the asymptotic expansion for large $-q^2$ in the
deep Euclidean region,
\begin{equation}\label{eq:Har}
  H_{-q^2/4} \approx -\log \frac{4}{-q^2} + \gamma + O(-q^2)^{-1}\,,
\end{equation}
the coefficient of the logarithm can be matched to that in the
parton-level calculation of the self-energy.  The {5D gauge coupling
constant in the AdS unit} is then determined to be
\begin{equation}\label{eq:match}
  g_5^2 =
      \frac{12\pi^2}{N_c}\, .
\end{equation}

\begin{table}[t!]
\begin{center}
\begin{tabular}{ |c||c|c|}
  \hline\hline \rule{0pt}{1.2em}
  ~ $U(1)_D\subset H$
  & embedding
  & $\rho$ representations\\ \hline\hline
  $Sp(N_c)$ &  $\begin{array}{c}SU(N_f)\times
                  U(1)_D \\ \subset Sp(2
                  N_f)\end{array}$ & $(\Ysymm\, ,
                                     +2)\oplus(\overline{\Ysymm}\,
                                     ,-2)\oplus({\rm
                                     adj},
                                     0)\oplus({\rm
                                     1},0)$ \\ \hline
  $SO(N_c)$ &  $\begin{array}{c}SU(N_f/2)\times
                  U(1)_D\\ \subset
                  SO(N_f)\end{array}$ & $(\Yasymm,
                                        +2) \oplus
                                        (\overline{\Yasymm},
                                        -2) \oplus
                                        ({\rm
                                        adj},
                                        0)\oplus{(\rm
                                        1},0)$ \\ \hline
  $SU(N_c)$ &  $\begin{array}{c}SU(N_1)\times
                  SU(N_2)\times U(1)_D\\ \subset
                  SU(N_f)\end{array}$ &
                                        $\begin{array}{c}(\Yfund, \overline{\Yfund}, 2) \oplus (\overline{\Yfund}, \Yfund, -2) \oplus \\  ({\rm adj}, 1, 0) \oplus (1, {\rm adj}, 0) \oplus (1, 1, 0)\end{array}$ \\ \hline
  \hline
\end{tabular}
\caption{Gauging the unbroken flavor symmetry in the AdS bulk, $H$, we
  can identify the spectrum of the $\rho$-mesons
  for the $Sp(N_c)$, $SO(N_c)$ and $SU(N_c)$ gauge theories. \label{tab:rho}}
\end{center}
\end{table}

The imaginary part of the the two-point function gives the decay width of the vector-mesons.  With the expressions above, one obtains a series of delta functions with no widths, which would not mimic the realistic situation.  In order to properly include widths of the KK gauge bosons, one needs to go beyond the leading approximation in the
$1/N_c$ expansion, which is beyond the current state of holographic QCD.  Here, we model the widths by the perturbative two-body decays into the lowest KK pion states (the dark matter).  The coupling $g_{\rho\pi\pi}$ is proportional to $g_5$, where the proportionality constant is given by the overlap integral of the relevant wave functions in the bulk, which are model-dependent.  As a proxy, we simply set $g_{\rho\pi\pi} = g_5$ which is supported by QCD data ($g_{\rho\pi\pi} = 5.95$ vs. $g_5=(12\pi^2/N_c)^{1/2} =6.28$~\cite{GarciaGudino:2012zz}), and obtain
\begin{equation}
  \Gamma_n^\rho =\frac{1}{4} D_R \frac{\beta^3}{96\pi} g_5^2 m_n,
  \qquad \beta=\sqrt{1-\frac{4m_\pi^2}{m_n^2}}\, .
  \label{eq:Gammarhon}
\end{equation}
Here, the factor $D_R = \Tr_R \,(T^a)^2$ (no sum over $a$) is the trace of the generator for the singlet $\rho$ vector on the representation $R$ of the pions (including their conjugates, see Table~\ref{tab:gauging}).  For the gauge groups considered here,
\begin{equation}
  D_R = \left\{ \begin{array}{lcc}
                  4(N_f-1)& & Sp(N_c)\,,\\
                  2(N_f+2) & & SO(N_c)\,,\\
                  4N_f  & & SU(N_c)\,.
                \end{array} \right.
              \label{eq:DR}
\end{equation}

The width can be included in the sum (Eq.~\ref{eq:PiHarmonic}) by modifying the external momentum
\begin{equation}\label{eq:qeff2}
  q^2 \rightarrow q_{\it eff}^2 = q^2 \left(1 + i \Gamma_n^\rho/m_n
  (m_n^2=q^2) \right)\,,
\end{equation}
for $q^2 > 4 m_\pi^2$.  Then the vacuum polarization function is
\begin{equation}
  \Pi(q^2)  = - \frac{1}{g_5^2}
  \sum_n \frac{2}{q_{\it eff}^2-4n+i\epsilon}\ .
\end{equation}
The inclusion of the width in this manner causes the asymptotic behavior of the harmonic number to change its
normalization at $O(g_5^2)$, which can be fixed by the rescaling
\begin{eqnarray}\label{eq:pi}
  \Pi(q^2)-\Pi(0)
  &=& - \frac{N_c}{24\pi^2}
      \left( 1- \frac{1}{\pi} \tan^{-1} \frac{1}{4} D_R
      \frac{g_5^2 }{96\pi} \right)^{-1}
      H_{-q_{\it eff}^2/4}\, .
\end{eqnarray}

Finally, to obtain the decay width for the dark photon $\Gamma_V$, the vacuum polarization needs to be normalized for the dark-photon current. The dark photon couples to the unnormalized generators
\begin{equation}
  Q=\left\{
  \begin{array}{lcl}
    {\rm diag}(\underbrace{+1, \cdots, +1}_{N_f},
    \underbrace{-1, \cdots, -1}_{N_f}  & &Sp(N_c) \\
      {\rm diag}(\underbrace{+1, \cdots, +1}_{N_f/2},
    \underbrace{-1, \cdots, -1}_{N_f/2})
                                                           & &SO(N_c)
    \end{array}
    \right.
\end{equation}
where $Q$ is the unnormalized generator of the dark photon.  Therefore the width of
the dark photon is given by
\begin{equation}
\label{2ptcorr}
  m_V \Gamma_V = a_V e_D^2 m_V^2 {\Im m} \Pi(m_V^2)\,,
\end{equation}
where $a_V$ is a normalization factor relating the 2-pt correlation function for the vector $\rho$-meson and hidden-photon
\begin{equation}\label{eq:aV}
  a_V  = \left\{ \begin{array}{lcc}
                  N_f& & Sp(N_c) \,,\\
                  N_f/2 & & SO(N_c)\,.\\
                \end{array} \right.
\end{equation}
This is the main result of this section, and Eq.~(\ref{2ptcorr}) will be used to model the width of the dark photon into dark-hadrons.

For $SU(N_c)$, the $SU(N_f)_L \times SU(N_f)_R \times U(1)_B$ symmetry is gauged in the AdS bulk.  The $\rho$-mesons belong to the unbroken $SU(N_f)_V$ subgroup, while $\omega$ belongs to the unbroken $U(1)_B$.
The vector boson masses of $\rho$ and $\omega$ are degenerate because they are both fixed only by the boundary conditions---independent of how the symmetry is broken in the bulk by the $X$ scalar---in good agreement with QCD.  The generator in the vector-vector correlation function is
\begin{equation}
  Q = \left(
    \begin{array}{cc}
      I_{N_1} & 0 \\
      0 & I_{N_2}
    \end{array} \right)
  =  \frac{N_1-N_2}{N_f} I_{N_f}  + \frac{2}{N_f} \left(
    \begin{array}{cc}
    N_2 \cdot   I_{N_1} & 0 \\
      0 &  N_1 \cdot I_{N_2}
    \end{array} \right)\,,
\end{equation}
separated into the $U(1)_B$ generator and a traceless $U(1)\subset SU(N_f)$ generator. Therefore the $\omega$ contribution to the vector-vector two-point function comes with the weight $(N_1-N_2)^2/N_f$, while the $\rho$ contribution has the weight $4 N_1 N_2/N_f$.  However, pions do not carry baryon number, and hence $\omega$ does not couple to pions at the `tree-level.'  At the quantum level, pions do carry baryon number due to the winding number of its soliton (Skyrmion) solution~\cite{Witten:1983tx}.  Therefore, $\omega$ decays into $3\pi$ via the Wess--Zumino--Witten terms, due to interactions of the form
\begin{equation}
  {\cal L}_{\it eff} \sim \frac{g_5^2 N_c}{f_\pi}
  \epsilon^{\kappa\lambda\mu\nu} \omega_{\kappa\lambda}
  \rho^a_{\mu\nu} \pi^a
  + \frac{g_5 N_c}{f_\pi^3} \epsilon^{\kappa\lambda\mu\nu}
  \omega_\kappa \partial_\lambda \pi^a \partial_\mu \pi^b \partial_\nu
  \pi^c f^{abc}\,.
\end{equation}
We obtain
\begin{eqnarray}\label{eq:piSUN}
  \Pi(q^2)-\Pi(0)
  &=& - \frac{N_c N_f}{12\pi^2} \left[ \frac{4 N_1 N_2}{N_f}
      \left( 1- \frac{1}{\pi} \tan^{-1}
      \frac{\Gamma_{\rho_\infty}}{m_{\rho_\infty}} \right)^{-1}
      H_{-q_{\it eff,\rho}^2/4} \right. \nonumber \\
  & & \left.
      + \frac{(N_1-N_2)^2}{N_f} \left( 1- \frac{1}{\pi}
      \tan^{-1}\frac{\Gamma_{\omega_\infty}}{m_{\omega_\infty}} \right)^{-1}
      H_{-q_{\it eff,\omega}^2/4} \right]\,. \nonumber \\
\end{eqnarray}
Here,
\begin{eqnarray}\label{eq:qeffSUN}
  q_{\it eff,\rho}^2 = q^2 \left( 1+i \frac{\Gamma_{\rho}}{m_{\rho}}(m_\rho^2=q^2) \right)\,, ~~~~~~~~~
  q_{\it eff,\omega}^2 =q^2 \left( 1+i
  \frac{\Gamma_{\omega}}{m_{\omega}}(m_\omega^2=q^2)
\right)\,.
\end{eqnarray}

In summary, the main results of the holographic QCD discussion are as follows: The invisible width of the dark photon into dark-hadrons of the strongly coupled SIMP sector can be modeled in $Sp(N_c)$ and $SO(N_c)$ theories using Eq.~\eqref{2ptcorr}, along with Eqs.~\eqref{eq:aV}, \eqref{eq:pi} and  \eqref{eq:qeff2} and in $SU(N_c)$ theories using Eqs.~\eqref{2ptcorr}, \eqref{eq:piSUN} and \eqref{eq:qeffSUN} and $a_V=1$.

{It would be interesting to study how the phenomenology varies for different gauge groups, $N_c$, $N_f$, and quark masses in a comprehensive way.  We leave this exploration to future studies.}

\bibliographystyle{JHEP}
\bibliography{biblio}{}

\providecommand{\href}[2]{#2}\begingroup\raggedright\begin{thebibliography}{10}

\bibitem{Hochberg:2014dra}
Y.~Hochberg, E.~Kuflik, T.~Volansky, and J.~G. Wacker, {\it {Mechanism for
  Thermal Relic Dark Matter of Strongly Interacting Massive Particles}},  {\em
  Phys. Rev. Lett.} {\bf 113} (2014) 171301,
  [\href{http://arxiv.org/abs/1402.5143}{{\tt arXiv:1402.5143}}].

\bibitem{Carlson:1992fn}
E.~D. Carlson, M.~E. Machacek, and L.~J. Hall, {\it {Self-interacting dark
  matter}},  {\em Astrophys. J.} {\bf 398} (1992) 43--52.

\bibitem{Hochberg:2014kqa}
Y.~Hochberg, E.~Kuflik, H.~Murayama, T.~Volansky, and J.~G. Wacker, {\it {Model
  for Thermal Relic Dark Matter of Strongly Interacting Massive Particles}},
  {\em Phys. Rev. Lett.} {\bf 115} (2015), no.~2 021301,
  [\href{http://arxiv.org/abs/1411.3727}{{\tt arXiv:1411.3727}}].

\bibitem{Wess:1971yu}
J.~Wess and B.~Zumino, {\it {Consequences of Anomalous Ward Identities}},  {\em
  Phys. Lett.} {\bf B37} (1971) 95.

\bibitem{Witten:1983tw}
E.~Witten, {\it {Global Aspects of Current Algebra}},  {\em Nucl. Phys.} {\bf
  B223} (1983) 422--432.

\bibitem{Witten:1983tx}
E.~Witten, {\it {Current Algebra, Baryons, and Quark Confinement}},  {\em Nucl.
  Phys.} {\bf B223} (1983) 433--444.

\bibitem{Choi:2015bya}
S.-M. Choi and H.~M. Lee, {\it {\uppercase{SIMP} Dark Matter with Gauged
  Z$_{3}$ Symmetry}},  {\em JHEP} {\bf 09} (2015) 063,
  [\href{http://arxiv.org/abs/1505.00960}{{\tt arXiv:1505.00960}}].

\bibitem{Lee:2015gsa}
H.~M. Lee and M.-S. Seo, {\it {Communication with \uppercase{SIMP} dark mesons
  via Z′ -portal}},  {\em Phys. Lett.} {\bf B748} (2015) 316--322,
  [\href{http://arxiv.org/abs/1504.00745}{{\tt arXiv:1504.00745}}].

\bibitem{Bernal:2015bla}
N.~Bernal, C.~Garc{\'\i a-}Cely, and R.~Rosenfeld, {\it {\uppercase{WIMP} and
  \uppercase{SIMP} Dark Matter from the Spontaneous Breaking of a Global
  Group}},  {\em JCAP} {\bf 1504} (2015), no.~04 012,
  [\href{http://arxiv.org/abs/1501.01973}{{\tt arXiv:1501.01973}}].

\bibitem{Schwaller:2015tja}
P.~Schwaller, {\it {Gravitational Waves from a Dark Phase Transition}},  {\em
  Phys. Rev. Lett.} {\bf 115} (2015), no.~18 181101,
  [\href{http://arxiv.org/abs/1504.07263}{{\tt arXiv:1504.07263}}].

\bibitem{Bernal:2015xba}
N.~Bernal and X.~Chu, {\it {$Z_2$ \uppercase{SIMP} Dark Matter}},
  \href{http://arxiv.org/abs/1510.08527}{{\tt arXiv:1510.08527}}.

\bibitem{Bernal:2015ova}
N.~Bernal, X.~Chu, C.~Garc{\'\i a-}Cely, T.~Hambye, and B.~Zaldivar, {\it
  {Production Regimes for Self-Interacting Dark Matter}},
  \href{http://arxiv.org/abs/1510.08063}{{\tt arXiv:1510.08063}}.

\bibitem{Hansen:2015yaa}
M.~Hansen, K.~Langæble, and F.~Sannino, {\it {\uppercase{SIMP} model at NNLO
  in chiral perturbation theory}},  {\em Phys. Rev.} {\bf D92} (2015), no.~7
  075036, [\href{http://arxiv.org/abs/1507.01590}{{\tt arXiv:1507.01590}}].

\bibitem{Clowe:2003tk}
D.~Clowe, A.~Gonzalez, and M.~Markevitch, {\it {Weak Lensing Mass
  Reconstruction of the Interacting Cluster 1E0657-558: Direct Evidence for the
  Existence of Dark Matter}},  {\em Astrophys. J.} {\bf 604} (2004) 596--603,
  [\href{http://arxiv.org/abs/astro-ph/0312273}{{\tt astro-ph/0312273}}].

\bibitem{Markevitch:2003at}
M.~Markevitch, A.~H. Gonzalez, D.~Clowe, A.~Vikhlinin, L.~David, W.~Forman,
  C.~Jones, S.~Murray, and W.~Tucker, {\it {Direct Constraints on the Dark
  Matter Self-Interaction Cross-Section from the Merging Galaxy Cluster
  1E0657-56}},  {\em Astrophys. J.} {\bf 606} (2004) 819--824,
  [\href{http://arxiv.org/abs/astro-ph/0309303}{{\tt astro-ph/0309303}}].

\bibitem{Randall:2007ph}
S.~W. Randall, M.~Markevitch, D.~Clowe, A.~H. Gonzalez, and M.~Bradac, {\it
  {Constraints on the Self-Interaction Cross-Section of Dark Matter from
  Numerical Simulations of the Merging Galaxy Cluster 1E 0657-56}},  {\em
  Astrophys. J.} {\bf 679} (2008) 1173--1180,
  [\href{http://arxiv.org/abs/0704.0261}{{\tt arXiv:0704.0261}}].

\bibitem{Rocha:2012jg}
M.~Rocha, A.~H.~G. Peter, J.~S. Bullock, M.~Kaplinghat, S.~Garrison-Kimmel,
  J.~Onorbe, and L.~A. Moustakas, {\it {Cosmological Simulations with
  Self-Interacting Dark Matter I: Constant Density Cores and Substructure}},
  {\em Mon. Not. Roy. Astron. Soc.} {\bf 430} (2013) 81--104,
  [\href{http://arxiv.org/abs/1208.3025}{{\tt arXiv:1208.3025}}].

\bibitem{Peter:2012jh}
A.~H.~G. Peter, M.~Rocha, J.~S. Bullock, and M.~Kaplinghat, {\it {Cosmological
  Simulations with Self-Interacting Dark Matter Ii: Halo Shapes Vs.
  Observations}},  {\em Mon. Not. Roy. Astron. Soc.} {\bf 430} (2013) 105,
  [\href{http://arxiv.org/abs/1208.3026}{{\tt arXiv:1208.3026}}].

\bibitem{Spergel:1999mh}
D.~N. Spergel and P.~J. Steinhardt, {\it {Observational Evidence for
  Selfinteracting Cold Dark Matter}},  {\em Phys. Rev. Lett.} {\bf 84} (2000)
  3760--3763, [\href{http://arxiv.org/abs/astro-ph/9909386}{{\tt
  astro-ph/9909386}}].

\bibitem{deBlok:2009sp}
W.~J.~G. de~Blok, {\it {The Core-Cusp Problem}},  {\em Adv. Astron.} {\bf 2010}
  (2010) 789293, [\href{http://arxiv.org/abs/0910.3538}{{\tt
  arXiv:0910.3538}}].

\bibitem{BoylanKolchin:2011de}
M.~Boylan-Kolchin, J.~S. Bullock, and M.~Kaplinghat, {\it {Too Big to Fail? the
  Puzzling Darkness of Massive Milky Way Subhaloes}},  {\em Mon. Not. Roy.
  Astron. Soc.} {\bf 415} (2011) L40,
  [\href{http://arxiv.org/abs/1103.0007}{{\tt arXiv:1103.0007}}].

\bibitem{Kaplinghat:2015aga}
M.~Kaplinghat, S.~Tulin, and H.-B. Yu, {\it {Dark Matter Halos as Particle
  Colliders: a Unified Solution to Small-Scale Structure Puzzles from Dwarfs to
  Clusters}},  \href{http://arxiv.org/abs/1508.03339}{{\tt arXiv:1508.03339}}.

\bibitem{Zavala:2012us}
J.~Zavala, M.~Vogelsberger, and M.~G. Walker, {\it {Constraining
  Self-Interacting Dark Matter with the Milky Way's Dwarf Spheroidals}},  {\em
  Monthly Notices of the Royal Astronomical Society: Letters} {\bf 431} (2013)
  L20--L24, [\href{http://arxiv.org/abs/1211.6426}{{\tt arXiv:1211.6426}}].

\bibitem{Vogelsberger:2012ku}
M.~Vogelsberger, J.~Zavala, and A.~Loeb, {\it {Subhaloes in Self-Interacting
  Galactic Dark Matter Haloes}},  {\em Mon. Not. Roy. Astron. Soc.} {\bf 423}
  (2012) 3740, [\href{http://arxiv.org/abs/1201.5892}{{\tt arXiv:1201.5892}}].

\bibitem{Massey:2015dkw}
R.~Massey et~al., {\it {The behaviour of dark matter associated with four
  bright cluster galaxies in the 10 kpc core of Abell 3827}},  {\em Mon. Not.
  Roy. Astron. Soc.} {\bf 449} (2015), no.~4 3393--3406,
  [\href{http://arxiv.org/abs/1504.03388}{{\tt arXiv:1504.03388}}].

\bibitem{Kahlhoefer:2015vua}
F.~Kahlhoefer, K.~Schmidt-Hoberg, J.~Kummer, and S.~Sarkar, {\it {On the
  Interpretation of Dark Matter Self-Interactions in Abell 3827}},  {\em Mon.
  Not. Roy. Astron. Soc.} {\bf 452} (2015), no.~1 L54--L58,
  [\href{http://arxiv.org/abs/1504.06576}{{\tt arXiv:1504.06576}}].

\bibitem{Hook:2010tw}
A.~Hook, E.~Izaguirre, and J.~G. Wacker, {\it {Model Independent Bounds on
  Kinetic Mixing}},  {\em Adv. High Energy Phys.} {\bf 2011} (2011) 859762,
  [\href{http://arxiv.org/abs/1006.0973}{{\tt arXiv:1006.0973}}].

\bibitem{Izaguirre:2013uxa}
E.~Izaguirre, G.~Krnjaic, P.~Schuster, and N.~Toro, {\it {New Electron
  Beam-Dump Experiments to Search for MeV to Few-GeV Dark Matter}},  {\em Phys.
  Rev.} {\bf D88} (2013) 114015, [\href{http://arxiv.org/abs/1307.6554}{{\tt
  arXiv:1307.6554}}].

\bibitem{Essig:2013vha}
R.~Essig, J.~Mardon, M.~Papucci, T.~Volansky, and Y.-M. Zhong, {\it
  {Constraining Light Dark Matter with Low-Energy $e^+e^-$ Colliders}},  {\em
  JHEP} {\bf 11} (2013) 167, [\href{http://arxiv.org/abs/1309.5084}{{\tt
  arXiv:1309.5084}}].

\bibitem{Curtin:2014cca}
D.~Curtin, R.~Essig, S.~Gori, and J.~Shelton, {\it {Illuminating Dark Photons
  with High-Energy Colliders}},  {\em JHEP} {\bf 02} (2015) 157,
  [\href{http://arxiv.org/abs/1412.0018}{{\tt arXiv:1412.0018}}].

\bibitem{Schael:2013ita}
{\bf DELPHI, OPAL, LEP Electroweak, ALEPH, L3} Collaboration, S.~Schael et~al.,
  {\it {Electroweak Measurements in Electron-Positron Collisions at
  W-Boson-Pair Energies at LEP}},  {\em Phys. Rept.} {\bf 532} (2013) 119--244,
  [\href{http://arxiv.org/abs/1302.3415}{{\tt arXiv:1302.3415}}].

\bibitem{Baer:2013cma}
H.~Baer, T.~Barklow, K.~Fujii, Y.~Gao, A.~Hoang, S.~Kanemura, J.~List, H.~E.
  Logan, A.~Nomerotski, M.~Perelstein, et~al., {\it {The International Linear
  Collider Technical Design Report - Volume 2: Physics}},
  \href{http://arxiv.org/abs/1306.6352}{{\tt arXiv:1306.6352}}.

\bibitem{Lees:2014xha}
{\bf BaBar} Collaboration, J.~P. Lees et~al., {\it {Search for a Dark Photon in
  $e^+e^-$ Collisions at BaBar}},  {\em Phys. Rev. Lett.} {\bf 113} (2014),
  no.~20 201801, [\href{http://arxiv.org/abs/1406.2980}{{\tt
  arXiv:1406.2980}}].

\bibitem{CMS:xwa}
{\bf CMS} Collaboration, {\it {Properties of the Higgs-like boson in the decay
  H to ZZ to 4l in pp collisions at $\sqrt{s} =7$ and 8 TeV,
  CMS-PAS-HIG-13-002}}, .

\bibitem{Chatrchyan:2013tia}
{\bf CMS} Collaboration, S.~Chatrchyan et~al., {\it {Measurement of the
  differential and double-differential Drell-Yan cross sections in
  proton-proton collisions at $\sqrt{s} =$ 7 TeV}},  {\em JHEP} {\bf 12} (2013)
  030, [\href{http://arxiv.org/abs/1310.7291}{{\tt arXiv:1310.7291}}].

\bibitem{Cline:2014dwa}
J.~M. Cline, G.~Dupuis, Z.~Liu, and W.~Xue, {\it {The Windows for Kinetically
  Mixed $Z'$-Mediated Dark Matter and the Galactic Center Gamma Ray Excess}},
  {\em JHEP} {\bf 08} (2014) 131, [\href{http://arxiv.org/abs/1405.7691}{{\tt
  arXiv:1405.7691}}].

\bibitem{ATLAS:2013jma}
{\bf ATLAS} Collaboration, {\it {Search for high-mass dilepton resonances in
  20~$fb^{-1}$ of $pp$ collisions at $\sqrt s = 8$~TeV with the ATLAS
  experiment, ATLAS-CONF-2013-017}}, .

\bibitem{Aubert:2008as}
{\bf BaBar} Collaboration, B.~Aubert et~al., {\it {Search for Invisible Decays
  of a Light Scalar in Radiative Transitions $\Upsilon(3S) \to \gamma A^0$}},
  in {\em {Proceedings, 34th International Conference on High Energy Physics
  (ICHEP 2008)}}, 2008.
\newblock \href{http://arxiv.org/abs/0808.0017}{{\tt arXiv:0808.0017}}.

\bibitem{Shoemaker:2011vi}
I.~M. Shoemaker and L.~Vecchi, {\it {Unitarity and Monojet Bounds on Models for
  DAMA, CoGeNT, and CRESST-II}},  {\em Phys. Rev.} {\bf D86} (2012) 015023,
  [\href{http://arxiv.org/abs/1112.5457}{{\tt arXiv:1112.5457}}].

\bibitem{Abdallah:2008aa}
{\bf DELPHI} Collaboration, J.~Abdallah et~al., {\it {Search for one large
  extra dimension with the DELPHI detector at LEP}},  {\em Eur. Phys. J.} {\bf
  C60} (2009) 17--23, [\href{http://arxiv.org/abs/0901.4486}{{\tt
  arXiv:0901.4486}}].

\bibitem{Adler:1969er}
S.~L. Adler and W.~A. Bardeen, {\it {Absence of Higher Order Corrections in the
  Anomalous Axial Vector Divergence Equation}},  {\em Phys. Rev.} {\bf 182}
  (1969) 1517--1536.

\bibitem{Kuflik:2015isi}
E.~Kuflik, M.~Perelstein, N.~R.-L. Lorier, and Y.-D. Tsai, {\it {Elastically
  Decoupling Dark Matter}},  \href{http://arxiv.org/abs/1512.04545}{{\tt
  arXiv:1512.04545}}.

\bibitem{Madhavacheril:2013cna}
M.~S. Madhavacheril, N.~Sehgal, and T.~R. Slatyer, {\it {Current Dark Matter
  Annihilation Constraints from CMB and Low-Redshift Data}},  {\em Phys. Rev.}
  {\bf D89} (2014) 103508, [\href{http://arxiv.org/abs/1310.3815}{{\tt
  arXiv:1310.3815}}].

\bibitem{Essig:2013goa}
R.~Essig, E.~Kuflik, S.~D. McDermott, T.~Volansky, and K.~M. Zurek, {\it
  {Constraining Light Dark Matter with Diffuse X-Ray and Gamma-Ray
  Observations}},  {\em JHEP} {\bf 11} (2013) 193,
  [\href{http://arxiv.org/abs/1309.4091}{{\tt arXiv:1309.4091}}].

\bibitem{Karch:2006pv}
A.~Karch, E.~Katz, D.~T. Son, and M.~A. Stephanov, {\it {Linear Confinement and
  AdS/QCD}},  {\em Phys. Rev.} {\bf D74} (2006) 015005,
  [\href{http://arxiv.org/abs/hep-ph/0602229}{{\tt hep-ph/0602229}}].

\bibitem{'tHooft:1973jz}
G.~'t~Hooft, {\it {A Planar Diagram Theory for Strong Interactions}},  {\em
  Nucl. Phys.} {\bf B72} (1974) 461.

\bibitem{Batell:2009yf}
B.~Batell, M.~Pospelov, and A.~Ritz, {\it {Probing a Secluded U(1) at
  B-Factories}},  {\em Phys. Rev.} {\bf D79} (2009) 115008,
  [\href{http://arxiv.org/abs/0903.0363}{{\tt arXiv:0903.0363}}].

\bibitem{Agashe:2014kda}
{\bf Particle Data Group} Collaboration, K.~A. Olive et~al., {\it {Review of
  Particle Physics}},  {\em Chin. Phys.} {\bf C38} (2014) 090001.

\bibitem{Strassler:2006im}
M.~J. Strassler and K.~M. Zurek, {\it {Echoes of a Hidden Valley at Hadron
  Colliders}},  {\em Phys. Lett.} {\bf B651} (2007) 374--379,
  [\href{http://arxiv.org/abs/hep-ph/0604261}{{\tt hep-ph/0604261}}].

\bibitem{Kang:2008ea}
J.~Kang and M.~A. Luty, {\it {Macroscopic Strings and `Quirks' at Colliders}},
  {\em JHEP} {\bf 11} (2009) 065, [\href{http://arxiv.org/abs/0805.4642}{{\tt
  arXiv:0805.4642}}].

\bibitem{Berends:1987zz}
F.~A. Berends, G.~J.~H. Burgers, C.~Mana, M.~Mart{\'\i ne}z, and W.~L. van
  Neerven, {\it {Radiative Corrections to the Process $e^+ e^-$ to Neutrino
  Anti-Neutrino $\gamma$}},  {\em Nucl. Phys.} {\bf B301} (1988) 583.

\bibitem{Higuchi}
C.~Hearty, T.~Higuchi, Y.~Iwasaki, T.~Iwashita, C.~Li, and K.~Miyabayashi.
  private communications.

\bibitem{Agnese:2013jaa}
{\bf SuperCDMS} Collaboration, R.~Agnese et~al., {\it {Search for Low-Mass
  Weakly Interacting Massive Particles Using Voltage-Assisted Calorimetric
  Ionization Detection in the SuperCDMS Experiment}},  {\em Phys. Rev. Lett.}
  {\bf 112} (2014), no.~4 041302, [\href{http://arxiv.org/abs/1309.3259}{{\tt
  arXiv:1309.3259}}].

\bibitem{Agnese:2014aze}
{\bf SuperCDMS} Collaboration, R.~Agnese et~al., {\it {Search for Low-Mass
  Weakly Interacting Massive Particles with SuperCDMS}},  {\em Phys. Rev.
  Lett.} {\bf 112} (2014), no.~24 241302,
  [\href{http://arxiv.org/abs/1402.7137}{{\tt arXiv:1402.7137}}].

\bibitem{Agnese:2015nto}
{\bf SuperCDMS} Collaboration, R.~Agnese et~al., {\it {New Results from the
  Search for Low-Mass Weakly Interacting Massive Particles with the CDMS Low
  Ionization Threshold Experiment}},  {\em Phys. Rev. Lett.} {\bf 116} (2016),
  no.~7 071301, [\href{http://arxiv.org/abs/1509.02448}{{\tt
  arXiv:1509.02448}}].

\bibitem{Essig:2011nj}
R.~Essig, J.~Mardon, and T.~Volansky, {\it {Direct Detection of Sub-GeV Dark
  Matter}},  {\em Phys. Rev.} {\bf D85} (2012) 076007,
  [\href{http://arxiv.org/abs/1108.5383}{{\tt arXiv:1108.5383}}].

\bibitem{Graham:2012su}
P.~W. Graham, D.~E. Kaplan, S.~Rajendran, and M.~T. Walters, {\it
  {Semiconductor Probes of Light Dark Matter}},  {\em Phys. Dark Univ.} {\bf 1}
  (2012) 32--49, [\href{http://arxiv.org/abs/1203.2531}{{\tt
  arXiv:1203.2531}}].

\bibitem{Hochberg:2015pha}
Y.~Hochberg, Y.~Zhao, and K.~M. Zurek, {\it {Superconducting Detectors for
  Super Light Dark Matter}},  \href{http://arxiv.org/abs/1504.07237}{{\tt
  arXiv:1504.07237}}.

\bibitem{Hochberg:2015fth}
Y.~Hochberg, M.~Pyle, Y.~Zhao, and K.~M. Zurek, {\it {Detecting Superlight Dark
  Matter with Fermi-Degenerate Materials}},
  \href{http://arxiv.org/abs/1512.04533}{{\tt arXiv:1512.04533}}.

\bibitem{Essig:2012yx}
R.~Essig, A.~Manalaysay, J.~Mardon, P.~Sorensen, and T.~Volansky, {\it {First
  Direct Detection Limits on Sub-GeV Dark Matter from \uppercase{XENON}10}},
  {\em Phys. Rev. Lett.} {\bf 109} (2012) 021301,
  [\href{http://arxiv.org/abs/1206.2644}{{\tt arXiv:1206.2644}}].

\bibitem{Angle:2011th}
{\bf XENON10} Collaboration, J.~Angle et~al., {\it {A search for light dark
  matter in \uppercase{XENON}10 data}},  {\em Phys. Rev. Lett.} {\bf 107}
  (2011) 051301, [\href{http://arxiv.org/abs/1104.3088}{{\tt
  arXiv:1104.3088}}]. [Erratum: Phys. Rev. Lett.110,249901(2013)].

\bibitem{Essig:2015cda}
R.~Essig, M.~Fernandez-Serra, J.~Mardon, A.~Soto, T.~Volansky, and T.-T. Yu,
  {\it {Direct Detection of Sub-Gev Dark Matter with Semiconductor Targets}},
  \href{http://arxiv.org/abs/1509.01598}{{\tt arXiv:1509.01598}}.

\bibitem{Bevan:2014iga}
{\bf Belle, BaBar} Collaboration, A.~J. Bevan et~al., {\it {The Physics of the
  B Factories}},  {\em Eur. Phys. J.} {\bf C74} (2014) 3026,
  [\href{http://arxiv.org/abs/1406.6311}{{\tt arXiv:1406.6311}}].

\bibitem{Fodor}
Z.~Fodor. private communications.

\bibitem{Soffer:2014ona}
A.~Soffer, {\it {Constraints on dark forces from the B factories and low-energy
  experiments}},  in {\em {Interplay between Particle and Astroparticle physics
  London, United Kingdom, August 18-22, 2014}}, 2014.
\newblock \href{http://arxiv.org/abs/1409.5263}{{\tt arXiv:1409.5263}}.

\bibitem{Alekhin:2015byh}
S.~Alekhin et~al., {\it {A facility to Search for Hidden Particles at the CERN
  SPS: the SHiP physics case}},  \href{http://arxiv.org/abs/1504.04855}{{\tt
  arXiv:1504.04855}}.

\bibitem{Titov}
A.~Chaus, J.~List, and M.~Titov, {\it Model-independent \uppercase{WIMP}
  searches at \uppercase{ILC} with single photon}, . a talk presented at
  International Workshop on Future Linear Colliders, Belgrade, 06--10 October
  2014,
  \url{http://agenda.linearcollider.org/event/6389/session/1/contribution/243/material/slides/1.pdf}.

\bibitem{Djouadi:2007ik}
{\bf ILC} Collaboration, G.~Aarons et~al., {\it {International Linear Collider
  Reference Design Report Volume 2: Physics at the ILC}},
  \href{http://arxiv.org/abs/0709.1893}{{\tt arXiv:0709.1893}}.

\bibitem{Riemann}
S.~Riemann, {\it Fermion pair production at a linear collider -- a sensitive
  tool for new physics searches}, .
  \url{http://tesla.desy.de/new_pages/TDR_CD/PartIII/references/LC-TH-2001-007.pdf}.

\bibitem{Essig:2010xa}
R.~Essig, P.~Schuster, N.~Toro, and B.~Wojtsekhowski, {\it {An Electron Fixed
  Target Experiment to Search for a New Vector Boson $A'$ Decaying to
  $e^+e^-$}},  {\em JHEP} {\bf 02} (2011) 009,
  [\href{http://arxiv.org/abs/1001.2557}{{\tt arXiv:1001.2557}}].

\bibitem{ArkaniHamed:2008qp}
N.~Arkani-Hamed and N.~Weiner, {\it {LHC Signals for a Superunified Theory of
  Dark Matter}},  {\em JHEP} {\bf 12} (2008) 104,
  [\href{http://arxiv.org/abs/0810.0714}{{\tt arXiv:0810.0714}}].

\bibitem{Baumgart:2009tn}
M.~Baumgart, C.~Cheung, J.~T. Ruderman, L.-T. Wang, and I.~Yavin, {\it
  {Non-Abelian Dark Sectors and Their Collider Signatures}},  {\em JHEP} {\bf
  04} (2009) 014, [\href{http://arxiv.org/abs/0901.0283}{{\tt
  arXiv:0901.0283}}].

\bibitem{Bai:2009it}
Y.~Bai and Z.~Han, {\it {Measuring the Dark Force at the LHC}},  {\em Phys.
  Rev. Lett.} {\bf 103} (2009) 051801,
  [\href{http://arxiv.org/abs/0902.0006}{{\tt arXiv:0902.0006}}].

\bibitem{Cheung:2009su}
C.~Cheung, J.~T. Ruderman, L.-T. Wang, and I.~Yavin, {\it {Lepton Jets in
  (Supersymmetric) Electroweak Processes}},  {\em JHEP} {\bf 04} (2010) 116,
  [\href{http://arxiv.org/abs/0909.0290}{{\tt arXiv:0909.0290}}].

\bibitem{Schwaller:2015gea}
P.~Schwaller, D.~Stolarski, and A.~Weiler, {\it {Emerging Jets}},  {\em JHEP}
  {\bf 05} (2015) 059, [\href{http://arxiv.org/abs/1502.05409}{{\tt
  arXiv:1502.05409}}].

\bibitem{Erlich:2014yha}
J.~Erlich, {\it {An Introduction to Holographic QCD for Nonspecialists}},  {\em
  Contemp. Phys.} {\bf 56} (2015) 159,
  [\href{http://arxiv.org/abs/1407.5002}{{\tt arXiv:1407.5002}}].

\bibitem{Csaki:2003dt}
C.~Csaki, C.~Grojean, H.~Murayama, L.~Pilo, and J.~Terning, {\it {Gauge
  Theories on an Interval: Unitarity without a Higgs}},  {\em Phys. Rev.} {\bf
  D69} (2004) 055006, [\href{http://arxiv.org/abs/hep-ph/0305237}{{\tt
  hep-ph/0305237}}].

\bibitem{Erlich:2005qh}
J.~Erlich, E.~Katz, D.~T. Son, and M.~A. Stephanov, {\it {QCD and a Holographic
  Model of Hadrons}},  {\em Phys. Rev. Lett.} {\bf 95} (2005) 261602,
  [\href{http://arxiv.org/abs/hep-ph/0501128}{{\tt hep-ph/0501128}}].

\bibitem{GarciaGudino:2012zz}
D.~Garcia~Gudino and G.~Toledo~Sanchez, {\it {The $\omega\rho\pi$ Coupling and
  the Influence of Heavier Resonances}},  {\em J. Phys. Conf. Ser.} {\bf 378}
  (2012) 012040.

\end{thebibliography}\endgroup

\end{document}